\documentclass[aps,pra,showpacs,twoside,10pt,floatfix,twocolumn,superscriptaddress]{revtex4-2}
\usepackage[colorlinks=true, citecolor=blue, urlcolor=blue]{hyperref}
\usepackage{epsfig,newlfont,amssymb,amsfonts,amsmath,bm,subfigure,palatino,mathtools,amsthm,braket,soul,enumitem,xcolor,graphics,graphicx,times,physics}
\usepackage[normalem]{ulem}
\begin{document}
\title{
Quantum-enhanced sensing from the interplay of long-range interactions and non-Hermiticity
}

\author{Keshav Das Agarwal}
\affiliation{Harish-Chandra Research Institute, Chhatnag Road, Jhunsi, Allahabad - 211019, India}
\affiliation{Homi Bhabha National Institute, Training School Complex, Anushakti Nagar, Mumbai 400 094, India}
\author{Tanoy Kanti Konar}
\affiliation{Harish-Chandra Research Institute, Chhatnag Road, Jhunsi, Allahabad - 211019, India}
\affiliation{Instytut Fizyki Teoretycznej, Wydzia\l{} Fizyki, Astronomii i Informatyki Stosowanej, Uniwersytet Jagiello\'nski, \L{}ojasiewicza 11, PL-30-348 Krak\'ow, Poland}	
\author{Leela Ganesh Chandra Lakkaraju}
\affiliation{Pitaevskii BEC Center, CNR-INO and Dipartimento di Fisica, Universit\`a di Trento, Via Sommarive 14, Trento, I-38123, Italy}	
\affiliation{INFN-TIFPA, Trento Institute for Fundamental Physics and Applications, Via Sommarive 14, Trento, I-38123, Italy}
\author{Aditi Sen(De)}
\affiliation{Harish-Chandra Research Institute,  Chhatnag Road, Jhunsi, Allahabad - 211019, India}
\affiliation{Homi Bhabha National Institute, Training School Complex, Anushakti Nagar, Mumbai 400 094, India}

\begin{abstract}
Long-range (LR) quantum spin systems offer promising advantages for quantum information processing and sensing. Here, we investigate parameter estimation in an long-range \(XX\) spin model coupled to a reservoir, which gives rise to an effective long-range \(\mathcal{RT}\)-symmetric non-Hermitian \(iXY\) Hamiltonian. The interactions extend up to a tunable coordination range and decay algebraically with distance, enabling a direct comparison between long-range and short-range (SR) regimes. Focusing on the estimation of the transverse magnetic field and anisotropy parameter, we initialize the system in a fully polarized state and analyze the resulting dynamical quantum Fisher information (QFI). We show that, with suitable tuning of the system parameters, both the time and system-size scaling of the QFI are enhanced in the LR regime relative to their SR counterparts. Moreover, the non-Hermitian LR model can exhibit superior dynamical QFI compared with the corresponding Hermitian model, demonstrating a genuine metrological advantage induced by the interplay of long-range interactions and non-Hermitian effects. In contrast, we establish a no-go result at the critical magnetic field: when the probe is prepared in the lowest-energy eigenstate, the QFI scaling remains identical for the Hermitian and non-Hermitian cases. 

\end{abstract}

\maketitle
% \section{to-do}
% Find the multiparamter QFI matrix for sensing $\gamma$ and $h$.

\section{Introduction}
\label{sec:intro}

Quantum sensing exploits quantum features to estimate physical parameters with precision beyond classical limits~\cite{Giovannetti2004, giovannetti_prl_2006, Giovannetti2011Apr, sensing_review_1, sensing_review_2, montenegro2024review, Agarwal2025_review}. Over the past decades, theoretical developments have demonstrated that collective many-body effects can substantially enhance sensitivity quantified through quantum Fisher information (QFI) \cite{Alushi2025}, particularly near critical points such as near gap-closing points~\cite{zanardi_1,zanardi_2,zanardi_3,mondal_prb_2022}, symmetry-breaking transitions~\cite{sarkar_prl_2022}, and localization-delocalization transition~\cite{sahoo_pra_2024}. 
%significant quantum advantages over classical strategies through the use of genuine quantum resources and processes. In particular, it has been shown that near gap-closing points~\cite{zanardi_1,zanardi_2,zanardi_3,mondal_prb_2022}, symmetry-breaking transitions~\cite{sarkar_prl_2022}, and localization-delocalization transition~\cite{sahoo_pra_2024} one can achieve enhanced precision, with 
%The quantum Fisher information (QFI) serving as a central figure of merit that quantifies the maximum accessible information about the parameter of interest.  
More precisely, when an interacting system of $N$ particles is employed as a probe, the QFI typically scales as $\sim N^{\mu}$, where $\mu$ characterizes the scaling exponent. The case of $\mu = 1$ corresponds to the standard quantum limit (SQL)~\cite{Helstrom1967, Holevo1973, caves_braunstein, Giovannetti2004}, which can be reached using conventional classical strategies. In contrast,  $\mu > 1$ exhibits a genuine quantum advantage with  $\mu \sim 2$ defining the Heisenberg limit (HL).
%, representing the optimal scaling achievable in standard quantum metrology. 
Interestingly, in certain quantum many-body interacting systems, even stronger than HL, referred to as super-Heisenberg limit, has been reported ~\cite{boixo_prl_2007, boixo_prl_2008, mishra_prl_2021, Yousefjani2023Oct, Mihailescu2024May, sahoo_pra_2024, Adani2024Aug, Mondal2024Jul, Mihailescu2024Jul, Yousefjani_pra_2025, Mihailescu2025Mar}, underscoring the potential of many-body resources for metrological gain. Beyond equilibrium encoding, in dynamical metrology,  the quantum Fisher information typically exhibits a joint scaling with system-size and evolution time of the form, \(N^{\mu} t^{\beta}\). In this framework, \(\mu>1\) signals a scaling beyond the standard quantum limit with respect to the \(N\), while \(\beta >2\) in the transient regime indicates enhanced dynamical encoding, thereby ensuring quantum advantage in this situation. It is worth noting, however, that this transient enhancement is fundamentally distinct from the asymptotic behavior -- at sufficiently long evolution times, the temporal QFI scales quadratically~\cite{rams_prx_2018,sahoo_pra_2024}.

In parallel, non-Hermitian physics has emerged as a powerful framework for sensing, particularly in open quantum systems where dissipation, monitoring, or post-selection induce effective non-Hermitian dynamics~\cite{Ueda_review,Bergholtz_rmp_2021,ElGanainy2018,luo_prl_2022}. Non-Hermitian lattice systems exhibiting the skin effect~\cite{budich_prl_2020,koch_2022_prr,Arandes2025} or nonreciprocal transport~\cite{McDonald2020Oct} have been identified as promising platforms for enhanced parameter estimation and sensing. A prominent early direction focused on exceptional points (EPs), where both eigenvalues and eigenvectors coalesce and the spectral response to weak perturbations becomes anomalously large~\cite{Wiersig_prl_2014,Wiersig_pra_2016}. At the same time, it is now well understood that an enhanced response near an EP does not automatically imply a corresponding enhancement in estimation precision, since mode nonorthogonality, excess noise, and more general information-theoretic constraints can compensate the apparent gain~\cite{wiersig_natcomm_2020,Lau2018Oct,ding_prl_2023}. These developments have motivated a broader view of non-Hermitian sensing, extending beyond fine-tuned spectral singularities toward dynamical and nonreciprocal protocols in which the sensing advantage arises from the structure of the effective non-Hermitian evolution itself~\cite{McDonald2020Oct,xiao_prl_2024}. Improved sensing based on non-Hermitian effects has been explored in a variety of classical or semiclassical platforms~\cite{chen_njp_2019} as well as in genuinely quantum systems~\cite{liu_prl_2016,Chen2017Aug,Lau2018Oct,McDonald2020Oct,chu_prl_2020}. More broadly, these developments challenge the conventional expectation that optimal sensing requires strictly Hermitian and coherent dynamics. Rather than offering a universal route to overcoming the ultimate sensitivity bounds of Hermitian sensors, effective non-Hermitian evolution provides a versatile setting in which transient dynamics, nonreciprocity, or robustness against technical noise can lead to meaningful operational advantages.

%A conventional expectation in quantum sensing is that suppressing environmental effects and maintaining Hermitian coherent dynamics should generically optimize metrological performance. Non-Hermitian systems, however, have repeatedly shown that this intuition is not universal. In particular,  effective non-Hermitian evolution, arising for instance through post-selected open-system trajectories, can enhance parameter sensitivity beyond what is achievable in corresponding Hermitian settings, which is also one of the main objectives of this current work. 
%Dynamical sensing in dissipative frameworks has been extensively explored to circumvent the limitations of static ground-state probes. By encoding the parameter of interest into the transient or steady-state dynamics of an open system, metrological advantages can be sustained even in the presence of environmental noise~\cite{Fernandez2017, Raghunandan2018, Hu2025, Vaidya2025, Zhang2026}. 

At the same time, long-range (LR) interacting quantum many-body systems, characterized by algebraically decaying coupling, 
%harboring long-range interactions 
have emerged as models with interesting and non-trivial physics~\cite{Campa2014, Vodola2014, Vodola2015, Maskara2022, Agarwal2026, lr_rmp_review} and have become experimentally accessible in 
%With experimental realizations of long-range interacting 
platforms, notably Rydberg atom arrays~\cite{Choi2016, Saffman2010}, polar molecule simulators~\cite{Carr2009, Lahaye2009} and trapped ions~\cite{HAFFNER2008, iontrap08, iontrap09, Britton_2012, iontrap_cirac04}, long-range models possess high entanglement~\cite{Koffel2012, Hauke_tagliocozzo_long_range_prl, anuradha2023production, ghosh2023entanglement}. The algebraically decaying couplings in such systems facilitate rapid entanglement generation and fast quantum scrambling~\cite{chen2019, tran2020, kuwahara2020}, accelerating the distribution of quantum information across the sensor and enabling scalable protocols that can dramatically surpass the standard quantum limit~\cite{Yousefjani2023Oct, monika2023better}. However, despite the individual successes of non-Hermitian enhanced sensitivity and long-range collective amplification, their combined role remains largely unexplored, particularly in the context of metrology. 
%. Specifically, a framework detailing the metrological potential of non-Hermitian sensors equipped with tunable long-range interactions is unexplored.

In this work, we address this gap by investigating the parameter estimation capabilities of a $\mathcal{RT}$-symmetric long-range non-Hermitian $iXY$ spin chain. By systematically tuning the interaction fall-off rates and the coordination range, we provide a systematic comparison of the scaling between the short- and long-range regimes for the dynamical encoding of a transverse magnetic field. We demonstrate that, in dynamical encoding protocols,  the interplay of non-Hermiticity and long-range couplings fundamentally alters the quantum Fisher information scaling with time and system-size. Remarkably, we report that the dynamical  QFI in the non-Hermitian long-range model can significantly surpass those of both the purely Hermitian long-range equivalent and the nearest-neighbor non-Hermitian models. We show that the $\mathcal{RT}$-symmetric broken regime provides a genuine metrological advantage, where the system-size scaling exponent of the QFI beats known Hermitian bounds. We also show that the non-Hermiticity gives amplification of QFI over its Hermitian counterpart, which persists in the long-range interactions, with greater amplification in the broken phase. Conversely, in the critical quantum metrology with a non-Hermitian model, we establish a definitive no-go result showing that long-range interactions do not improve the scaling beyond nearest-neighbor bounds. These findings identify dynamical non-Hermitian long-range systems as a promising route for achieving enhanced precision in quantum sensing.

%in the static sensing regime: when parameters are encoded in the steady state at critical magnetic fields, long-range interactions offer no improvement over the nearest-neighbor scaling bounds.

The paper is organized as follows. In Sec. \ref{sec:model}, we first introduce the long-range $iXY$ model, and its derivation from the open system set-up (Sec. \ref{subsec:effectiveH}), and finally the dynamical protocol for estimating the magnetic field via computing QFI (Sec. \ref{subsec:computQFId}). We describe the QFI scaling in long-range systems with both time and system-size in Sec. \ref{sec:sensing_advantage_dynamics}. The comparison between non-Hermitian and its Hermitian counterpart is presented in Sec. \ref{sec:nH_by_H}. The no-go result showing no benefit of LR over NN model is shown in Sec. \ref{sec:disadvantage_equilibrium}. We finally conclude in Sec. \ref{sec:conclusion}.

\section{Long-range $iXY$ model: dynamics and quantum Fisher information}
\label{sec:model}

We consider a one-dimensional long-range (LR) non-Hermitian \(iXY\) model consisting of \(N\) spin-\(1/2\) particles with algebraically decaying interactions. The Hamiltonian is given by
%~\cite{agarwal2025, agarwal2022detecting}
\begin{align}
    \nonumber H^{iLR} &= \sum_{j=1}^N \sum_{r=1}^{\mathcal{Z}} J_r(\alpha)\left[ \frac{(1+i\gamma)}{4} \sigma_j^x Z_{j+1}^{j+r-1} \sigma_{j+r}^x  \right.\\ 
    &\left. + \frac{(1-i\gamma)}{4} \sigma_j^y Z_{j+1}^{j+r-1} \sigma_{j+r}^y \right] + \frac{h}{2} \sum_{j=1}^{N}\sigma_j^z,
    \label{eq:Hamil}
\end{align}
where \(\sigma_j^i\) (\(i=x,y,z\)) denote the Pauli operators at site \(j\), and \(Z_{a}^{b}=\prod_{p=a}^{b}\sigma_p^z\) is a \((b\!-\!a\!+\!1)\)-body interactions. The coupling strengths decay algebraically with distance as \(J_{r}(\alpha)=\frac{1}{\mathcal{K}(\alpha) r^\alpha}\). The normalization is ensured by the Kac factor~\cite{kac_normalization}, defined as \(\mathcal{K}(\alpha)=\sum_{r=1}^{{\mathcal{Z}}} r^{-\alpha}\), which guarantees \(\sum_{r=1}^\mathcal{Z} J_r(\alpha)=1\) for all values of the decay exponent \(\alpha\), particularly important in the regime \(\alpha<1\). Here, \(\mathcal{Z}\) sets the cutoff of the interaction range, while \(\alpha\) controls how rapidly the interactions decay. The parameter \(\gamma\) characterizes the degree of non-Hermiticity, and \(h\) denotes the strength of the transverse magnetic field. Throughout this work, periodic boundary conditions are assumed, i.e., \(N+j \equiv j\).

The system possess the $\mathcal{RT}$-symmetry with $\mathcal{R}=\bigotimes_{k=1}^N \exp[-i\frac{\pi}{4}\sigma^z_k]$ as the rotation around the $z$-axis by $\pi/4$ angle ($\sigma^x \xrightarrow{\mathcal{R}} \sigma^y, \sigma^y \xrightarrow{\mathcal{R}} -\sigma^x$ and $\sigma^z \xrightarrow{\mathcal{R}} \sigma^z$) and $\mathcal{T}i\mathcal{T}=-i$ as the time-reversal operation acting as the complex conjugation. Therefore, the system described by the Hamiltonian in Eq.~(\ref{eq:Hamil}) follows $[H^{iLR}, \mathcal{RT}]=0$, i.e.,  $H^{iLR}\mathcal{R}=\mathcal{R}(H^{iLR})^*$. The parameters where the all the eigenstates of the system exhibits the $\mathcal{RT}$-symmetry, are in the unbroken phase (all eigenvalues are real), whereas when the eigenstates break the $\mathcal{RT}$-symmetry, the system is in the broken phase, having complex conjugate eigenvalue pairs. The transition between these two phases occurs at the exceptional points, where eigenvalues and eigenvectors simultaneously coaslesce.  
%are the set of parameter points of the Hamiltonian, separating these two phases.

Under the Jordan–Wigner followed by Fourier transformations (see Appendix~\ref{app:jw_ft}), the model 
%decomposes 
decouples into independent momentum sectors, $H^{iLR}=\bigoplus_{p=1}^{N/2-1}\mathcal{H}_p$, with each block acting on the reduced Nambu basis $\{\ket{0}_p, c_p^\dagger c_{-p}^\dagger\ket{0}_p\}$\cite{nambu1960, barouch_pra_1970, barouch_pra_1971, LSM_main, santoro_ising_beginners_2020}, where  $c_p^\dagger$ and $c_p$ denote fermionic creation and annihilation operators in the momentum space. In this representation, the system reduces to the effective two-level structure, given by
%takes the form
\begin{equation}
    \mathcal{H}_p=\left[\begin{array}{cc}
-h-J_p^{(R)} & -\gamma J_p^{(I)} \\
\gamma J_p^{(I)} & h+J_p^{(R)}
\end{array}\right],
\label{eq:ksea_P}
\end{equation}
where   $J_p (\alpha, {\mathcal{Z}}) = \sum_{r=1}^{\mathcal{Z}} J_r(\alpha) e^{ir\phi_p} \equiv J_p^{(R)}+iJ_p^{(I)}$ and $\phi_p=(2p-1)\pi/N$ labels the discrete quasi-momenta in the even-parity sector, which hosts the ground state for finite system sizes as well as the relevant dynamical subspace. The spectrum of each block is given by $\pm \epsilon(\phi_p)$, with
\begin{equation}
    \epsilon(\phi_p) =\sqrt{(h+J_p^{(R)})^2 -\gamma^2 J_p^{(I)}}.
    \label{eq:dispersion}
\end{equation}

The exceptional structure is determined by a different spectral criterion. The unbroken phase is characterized by the $\epsilon^2(\phi_p)>0$ for all momenta $\phi_p\in[0,\pi]$ while  $\epsilon^2(\phi_p)<0$ for some momentum $\phi_p$ signals the broken phase, given sets of imaginary $\epsilon(\phi_p)$~\cite{ganesh_aditi_factorization_surface,Agarwal2023May}. The exceptional line is determined by $\epsilon(\phi_p)=0$ and $\frac{d\epsilon(\phi_p)}{d\phi_p}=0$ for some $\phi_p$. We define the exceptional point $h_{e}$ as the smallest value of the magnetic field for which the spectrum becomes entirely real, i.e., for which $\epsilon(\phi_p)\in\mathbb{R}$ for all momentum sectors. For generic long-range couplings, this condition does not lead to a closed-form expression. We therefore determine $h_{e}$ numerically by scanning the magnetic-field parameter and monitoring the spectrum of all momentum blocks. The resulting value is obtained with a numerical precision of $10^{-9}$ for each choice of $\alpha$, $\mathcal{Z}$, and $\gamma$ considered in this work (see Appendix~\ref{app:ep_points} for the $h_e$ values with a fixed $\gamma=0.5$  and various $(\alpha, \mathcal{Z})$-values).

\subsection{Effective non-Hermitian long-range Hamiltonian}
\label{subsec:effectiveH}
First, we note that a Hermitian~\cite{Vodola2014, Vodola2015, Maity2019, Lakkaraju2022, lr_rmp_review} long-range extended \(XY\) model resembles the Hamiltonian in Eq.~(\ref{eq:Hamil}), although its physical origin is fundamentally different from the Hermitian ones. We now outline how such an effective non-Hermitian Hamiltonian naturally emerges from coupling the system to an external environment. To this end, we begin with a long-range \((\mathcal{Z}\!+\!1)\)-body interacting \(XX\) model described by
\begin{eqnarray}
    H_S^{XX} &=& \sum_{j=1}^N \sum_{r=1}^{\mathcal{Z}} \frac{J_r(\alpha)}{4} \left( \sigma_j^x Z_{j+1}^{j+r-1} \sigma_{j+r}^x + \sigma_j^y Z_{j+1}^{j+r-1} \sigma_{j+r}^y \right) \nonumber \\
    &+& \frac{h}{2} \sum_{j=1}^{N}\sigma_j^z,
    \label{eq:long_range_xx}
\end{eqnarray}
where \(Z_{j+1}^{j+r-1} = \prod_{p=j+1}^{j+r-1}\sigma_p^z\) denotes the \((r\!-\!1)\)-body interaction with decaying strength. We now assume that each spin is coupled to an environment, and the resulting open-system dynamics are governed by the Gorini-Kossakowski-Lindblad-Sudarshan (GKLS) master equation, given as
\begin{equation}
    \frac{d\rho}{dt} = -i [H_S, \rho] + \sum_{k} \gamma_k \mathcal{L}[A_k](\rho),
\end{equation}
where \(A_k\) are (possibly correlated) Lindblad jump operators, and \(\mathcal{L}[A_k](\rho) = A_k \rho A_k^\dagger - \frac{1}{2} \{A_k^\dagger A_k, \rho\}\) denotes the corresponding dissipator with rate \(\gamma_k\). Now one can introduce an effective Hamiltonian by ignoring the jump operation, given as \(H_{\text{eff}} = H_S - \frac{i}{2} \sum_k \gamma_k A_k^\dagger A_k,\) and the GKLS equation can be rewritten as
\begin{equation}
    \frac{d\rho}{dt} = -i\left(H_{\text{eff}} \rho - \rho H_{\text{eff}}^\dagger\right) + \sum_k \gamma_k A_k \rho A_k^\dagger.
\end{equation}
Upon post-selecting trajectories with no quantum jumps (i.e., continuous monitoring with no detection events), the recycling term vanishes, \(A_k \rho A_k^\dagger \equiv 0\), leading to a deterministic non-unitary evolution governed solely by \(H_{\text{eff}}\)~\cite{Zhang2024,Agarwal2023May}. The Hermitian contribution corresponds to the standard long-range \(XX\) model in a transverse field, given in Eq.~(\ref{eq:long_range_xx}). To engineer the desired LR non-Hermitian terms, we introduce non-local jump operators acting on pairs of sites \(j\) and \(j+r\), dressed by the Jordan-Wigner string, as
\begin{equation}
    A_{j,r} = \left( \sigma_j^- - Z_{j+1}^{j+r-1} \sigma_{j+r}^+ \right),\quad \gamma_{j,r} =  \sqrt{\gamma J_r(\alpha)}.
\end{equation}
This construction yields
\begin{align}
    \gamma_{j,r}A_{j,r}^\dagger A_{j,r} = & \gamma J_r(\alpha) \Big( \sigma_j^+ \sigma_j^- + \sigma_{j+r}^- \sigma_{j+r}^+ \nonumber \\
    & \left. - \sigma_j^+ Z_{j+1}^{j+r-1} \sigma_{j+r}^+ - \sigma_j^- Z_{j+1}^{j+r-1} \sigma_{j+r}^- \right).
\end{align}
Using the identities \(\sigma^{\pm}\sigma^{\mp} = \frac{1}{2}(1 \pm \sigma^z)\), we obtain
%\begin{widetext}
\begin{eqnarray}
    %\begin{aligned}
   && - \frac{i}{2} \sum_{j,r} \gamma_{j,r} A_{j,r}^\dagger A_{j,r} \nonumber\\
    &&=  - i \sum_{j=1}^N \sum_{r=1}^{\mathcal{Z}} \frac{\gamma J_r(\alpha)}{2} \left[ 1 + \frac{1}{2}(\sigma_j^z - \sigma_{j+r}^z) \right]  \nonumber\\
    &&\nonumber +  i \sum_{j=1}^N \sum_{r=1}^{\mathcal{Z}} \frac{\gamma J_r(\alpha)}{2} \left( \sigma_j^+ Z_{j+1}^{j+r-1} \sigma_{j+r}^+ + \sigma_j^- Z_{j+1}^{j+r-1} \sigma_{j+r}^- \right).\\
%\end{aligned}
\label{eq:nH_term}
\end{eqnarray}

%\end{widetext}
Since, under periodic boundary conditions, the sum over local magnetization differences vanishes, i.e., \(\sum_{j=1}^N (\sigma_j^z - \sigma_{j+r}^z) = 0\), the remaining constant contribution, \(-i \frac{N\gamma}{2} \mathbb{I}\), corresponds to a global imaginary energy shift that induces a uniform exponential decay of the wavefunction, \(\ket{\Psi(t)} \to e^{-\frac{N\gamma}{2}t} \ket{\Psi(t)}\). This factor is removed by normalization and does not affect the dynamics. Consequently, the first term in Eq.~(\ref{eq:nH_term}) can be discarded, while the second term generates the desired long-range non-Hermitian interactions. The resulting dynamics are thus governed by an \(\mathcal{RT}\)-symmetric effective Hamiltonian, given in Eq. (\ref{eq:Hamil}).
% \begin{equation}
%     H^{i\mathrm{LR}} = H_S - \frac{i}{2} \sum_{j,r} \gamma_{j,r} A_{j,r}^\dagger A_{j,r},
% \end{equation}
% which reproduces Eq.~(\ref{eq:Hamil}).

\subsection{Dynamical protocol to compute quantum Fisher information}
\label{subsec:computQFId}

%The resulting non-Hermitian dynamics provides an effective description of an open quantum system in the no-click post-selected trajectory. 
Starting from a fully polarized initial state $\ket{\Psi(0)}=\ket{0}^{\otimes N}=\bigoplus_{p=1}^{N/2-1}\ket{0}_p$, the time evolution remains factorized across momentum sectors, $\ket{\Psi(t)}=\bigoplus_{p=1}^{N/2-1}\ket{\psi(t)}_p$, with each mode evolving as $\ket{\psi(t)}_p = \mathcal{N}_t^{-1/2} U_p(t)\ket{\psi(0)}_p$, where $\mathcal{N}_t$ ensures state normalization at each time due to the non-unitary nature of the evolution. The effective propagator within each block reads
\begin{equation}
    U_p(t) = e^{-i\mathcal{H}_p t} = \cos[\epsilon(\phi_p) t]\, \mathbb{I}_p + i\frac{\sin[\epsilon(\phi_p) t]}{\epsilon(\phi_p)} \mathcal{H}_p,
\end{equation}
reflecting the non-Hermitian dynamics governed by \(\mathcal{H}_p\).  This dynamical protocol encodes Hamiltonian parameters (eg. \(h\), \(\gamma\) and \(\alpha\)) into the evolving quantum state $\ket{\Psi(t)}$, thereby defining a quantum metrological probe. For pure states, the sensitivity of this encoding is quantified via the QFI associated with a parameter $\theta$, given by 
\begin{align}
    &\mathcal{F}_\theta(\ket{\Psi}) 
    = 4 \ \mathrm{Re}\!\!\left[\langle \partial_\theta \Psi|\partial_\theta \Psi \rangle - |\langle \Psi|\partial_\theta \Psi \rangle|^2\right] \nonumber \\
    &=\! \sum_{\mathclap{p=1}}^{\mathclap{N/2-1}} 4 \ \mathrm{Re}\!\!\left[\langle \partial_\theta \psi_p(t)|\partial_\theta \psi_p(t) \rangle \!-\! |\langle \psi_p(t)|\partial_\theta \psi_p(t) \rangle|^2\right],
    \label{eq:qfi_theta}
\end{align}
where the additivity over momentum sectors follows directly from the factorized structure of the state, while the parameter dependence is inherited through the non-Hermitian spectral deformation of each $\mathcal{H}_p$ block.

In general, for an unbiased estimator $\hat{\theta}$ constructed from $m$ independent probes, the variance satisfies the quantum Cramér–Rao bound $\delta^2 \hat{\theta}\geq (m\mathcal{F}_\theta)^{-1}$ \cite{caves_braunstein}. The scaling of $\mathcal{F}_\theta$ with system-size $N$ and evolution time $t$, typically of the form $\mathcal{F}_\theta \sim N^\mu t^\beta$, with \(\mu\) and \(\beta\) being the scaling exponents which  determine the attainable precision enhancement of the protocol. In interaction-driven encoding schemes of the form $H=H_1+\theta H_2$, where $H_2$ contains $k$-body terms, the achievable scaling is constrained by $\mu \leq k$~\cite{Puig2025}, while dynamical amplification generically imposes $\beta \leq 2$ in the long-time regime. In the dynamical setting, observing scaling exponents \(\mu >k\) and \(\beta>2\) signifies a clear departure from standard quantum limit (SQL) behavior, indicating a genuine metrological advantage enabled by the underlying dynamics. Our aim here is to determine how these scalings, capturing the metrological benefit, are modified by the interplay between long-range interactions and non-Hermitian dynamics.

%Within the present framework, we restrict attention to pure-state probes, for which the QFI admits the compact form
% %\begin{align}
%     &\mathcal{F}_\theta(\ket{\Psi}) 
%     = 4 \mathrm{Re}\left(\langle \partial_\theta \Psi|\partial_\theta \Psi \rangle - |\langle \Psi|\partial_\theta \Psi \rangle|^2\right) \nonumber \\
%     &= \sum_{p=1}^{N/2-1} 4 \mathrm{Re}\left(\langle \partial_\theta \psi_p(t)|\partial_\theta \psi_p(t) \rangle - |\langle \psi_p(t)|\partial_\theta \psi_p(t) \rangle|^2\right),
% \end{align}

\section{Sensing advantage under long-range non-Hermitian interaction}
\label{sec:sensing_advantage_dynamics}

%In this work, we consider the magnetic field sensing, i.e., $\theta\equiv h$ in the long-range $iXY$ model and study the quantum Fisher information and its scaling with the resources, namely system-size $N$ and the time $t$ in the dynamical encoding with the non-Hermitian model $H^{iLR}$.

We now investigate the estimation of the transverse magnetic field, i.e., $\theta\equiv h$, in the long-range non-Hermitian \(iXY\) model by analyzing the quantum Fisher information and its scaling with the relevant resources, namely the system-size \(N\) and the evolution time \(t\). We focus on the transient (short- to intermediate-time) regime, where dynamical encoding under the non-unitary evolution generated by \(H^{iLR}\)
 can lead to enhanced sensitivity. In particular, we examine how the non-Hermitian nature of the model influences the metrological scaling and whether distinct behaviours emerge across the unbroken and broken phases. This analysis allows us to identify parameter regimes where non-Hermitian effects can enhance the achievable precision, which we will also address in the next section by comparing results obtained here with their Hermitian counterparts. A similar study can also be performed for estimating anisotropic parameters in dynamics (see Appendix \ref{app:gamma_sensing}). 

It is important to stress at the outset that the advantage studied here is not based on tuning the dynamics to an exceptional point. A large part of the current non-Hermitian sensing literature has focused on exceptional-point-based proposals, where the enhanced spectral response is often offset by excess noise, mode nonorthogonality, or stability limitations \cite{wiersig_natcomm_2020,wiersig_pra_2020,ding_prl_2023}. By contrast, the long-range \(iXY\) model studied here exhibits an enhanced \emph{dynamical} encoding rate over extended parameter regions, including both completely real and partially complex spectral sectors. Our results therefore belong to the broader class of non-Hermitian sensing protocols in which the metrological gain arises from non-unitary many-body dynamics rather than from a fine-tuned degeneracy \cite{chu_prl_2020,xiao_prl_2024,yang2026practical}.

\begin{figure}
    \includegraphics[width=\linewidth]{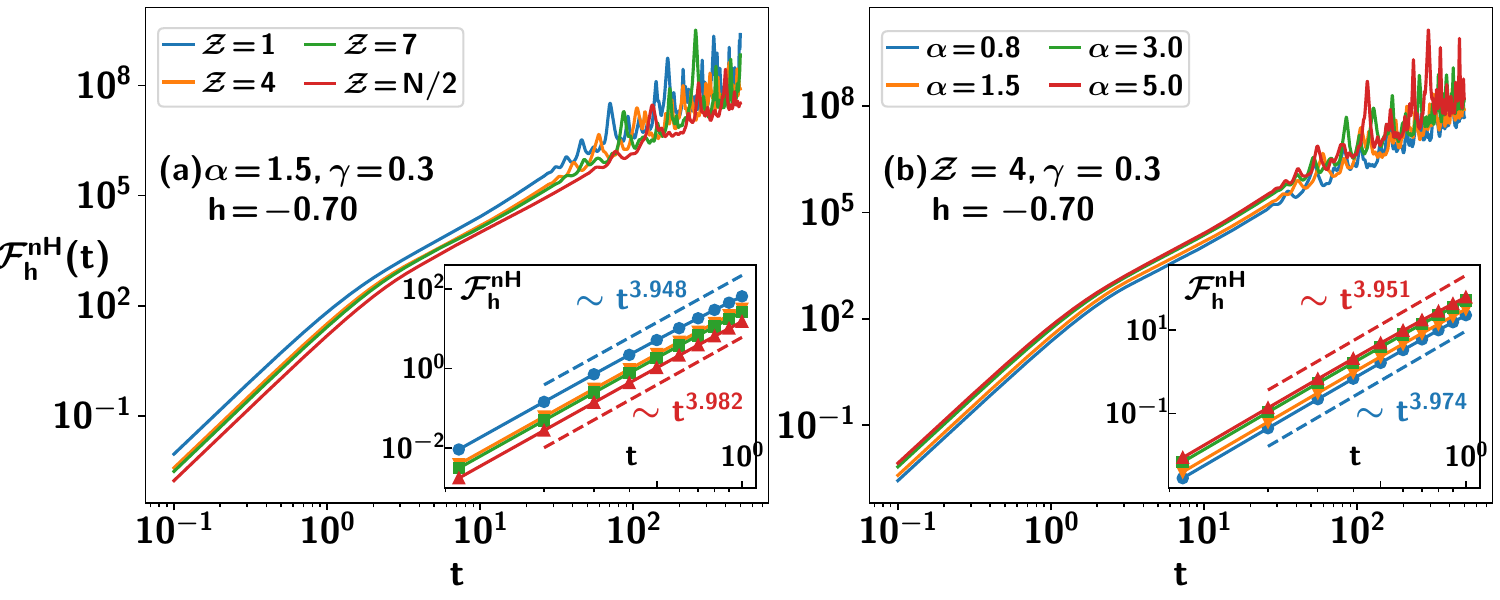}
    \caption{QFI of the dynamical state for sensing magnetic fields in the long-range $iXY$ model $H^{iLR}$. (a) QFI $\mathcal{F}_h^{nH}$ (ordinate) vs time $t$ (abscissa) for different coordination numbers \(\mathcal{Z}\) at fall-off rate $\alpha=1.5$.  (b) $\mathcal{F}_h^{nH}$ aganist \(t\) for different fall-off rates $\alpha$ with a fixed \(\mathcal{Z} =4\).  Both the insets show the scaling of QFI with time in the transient time $t\lesssim 2.0$. Here, the imaginary anisotropy is taken to be $\gamma=0.3$,  \(h =-0.70\) and the system-size is $N=1024$. All axes are dimensionless.}
    \label{fig:qfi_lr}
\end{figure}

\begin{figure}
    \includegraphics[width=\linewidth]{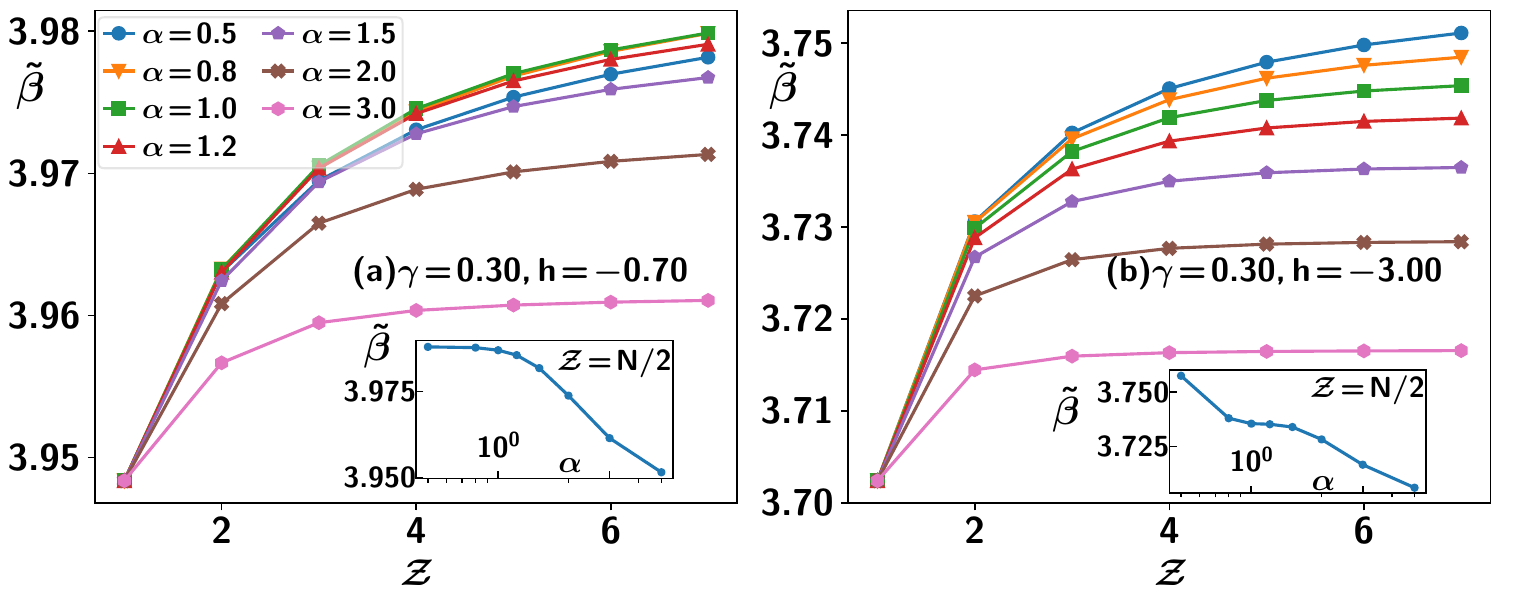}
    \caption{Scaling exponent $\tilde{\beta}$ (ordinate) of QFI, i.e.,  $\mathcal{F}_{h}^{nH}(t)\sim t^{\tilde{\beta}}$ in the transient time in the (a) broken and (b) unbroken regime against the coordination number $\mathcal{Z}$ (abscissa) for various fall-off rates $\alpha$. The insets show the scaling exponent for the $\mathcal{Z}=N/2$-case. Note that the long-range interactions increase $\tilde{\beta}$ with increasing coordination number and decreasing values of $\alpha$.
    %Here the imaginary anisotropy is $\gamma=0.3$ and the system-size is $N=1024$. 
    All other specifications are the same as in Fig. \ref{fig:qfi_lr}, and all axes are dimensionless.}
    \label{fig:transient_t_scaling}
\end{figure}

\subsection{Scaling of QFI with time -- Beating SQL with non-Hermiticity}

In the dynamical sensing protocol, the time-dependence of QFI reflects how efficiently the parameter is encoded into the evolving probe state. Typically, an increase of \(\mathcal{F}_h^{nH}(t)\) with time corresponds to the improved estimation precision (see Fig. \ref{fig:qfi_lr}). 
%scaling of QFI with time reveals the parameter encoding ability of the system on a given quantum probe, with improving precision as time increases. 
%Therefore, the high scaling of QFI with time sheds light on the quantum benefit of the system. 
While at large times, dynamical protocols are known to exhibit quadratic scaling, i.e., $\mathcal{F}_h\sim t^2$, it has been established that in the transient regime, the time scaling of QFI performs better as compared to the steady state regime \cite{plenio_time_sensing}. Here, we go beyond these conventional settings by investigating the role of long-range interactions in a non-Hermitian \(iXY\) model for magnetic-field estimation. Importantly, our analysis demonstrates the impact of non-Hermiticity and LR coupling in quantum sensing performance. 
%Here we study the benefit of the long-range interactions in the non-Hermitian model in the scaling of QFI for magnetic field sensing with time.

With the initial product state, the QFI $\mathcal{F}_h^{nH}$ increases polynomially with time, with distinct scaling exponents in the transient time ($t\lesssim 2$), and in the long time ($t\gg 100$) scenarios for all system-size. We first notice that the role of long-range interactions on QFI depends crucially on both the coordination number $\mathcal{Z}$ and the fall-off rate $\alpha$, as shown in Fig.~\ref{fig:qfi_lr}. Before discussing their impacts, let us fix that in the transient time, i.e., $t\lesssim2.0$,  $\mathcal{F}_{h}^{nH}\sim t^{\tilde{\beta}}$   and $\mathcal{F}_{h}^{nH}\sim t^\beta$ at long times, i.e., $t\gg 200$ where $\tilde{\beta}$ and $\beta$ denote the corresponding exponents.
%scaling exponents in the transient and long times respectively. 
For the nearest-neighbor (NN) case ($\mathcal{Z}=1$), the system has no $\alpha$-dependence in Eq.~(\ref{eq:Hamil}) and one obtains $\mathcal{F}_h^{nH}\sim t^{\tilde{\beta}_1}$, with $\tilde{\beta}_1\lesssim 4$ \cite{agarwal2025}.

{\it Effect of coordination number.} For a fixed decay exponent \(\alpha\), increasing the interaction range via the coordination number $\mathcal{Z}$  leads to a systematic enhancement of the transient-time scaling of the QFI (see Figs.~\ref{fig:qfi_lr}(a), and \ref{fig:transient_t_scaling}).  This demonstrates that long-range (LR) interactions provide a clear metrological advantage over the nearest-neighbor limit. Quantitatively, for $\alpha=1.5$, we find that  $\tilde{\beta}$ increases from $\tilde{\beta}_1=3.948$ to $\tilde{\beta}(\mathcal{Z}=N/2, \alpha=1.5)=3.982$, highlighting the benefit of extending the interaction range. It is important to note that the coordination number \(\mathcal{Z}\) can be directly engineered via multi-body interaction in several experimental platforms such as trapped ions \cite{Trapped_ion_nat_phys_ion_trapp_three_body_four_body_2023_monroe}, Rydberg atom arrays \cite{Zoller_three_body_cold_polar_molecules_2007}, optical lattices \cite{maciej_cirac_optical_lattices,daley_optical_lattice_2014} bosons in free space \cite{bosons_in_free_space_2014}, and cavity-mediated spin systems, circuit-QED systems \cite{ZZZ_superconducting_prl,zzz_interaction_superconducting_2024_thesis,circuit_qed_three_atoms}, NMR quantum simulation \cite{NMR_quantum_simulation}. It is important to note that Floquet engineering \cite{Floquet_review} can also lead to the multi-body interaction, which has also been implemented experimentally~\cite{floquet_experiment}; however, this might lead to unwanted heating during large time dynamics. Hence, our results provide a viable, physically accessible route to enhance sensitivity in quantum magnetometry. 
%The coordination number $\mathcal{Z}$ increases the QFI $\mathcal{F}_h^{nH}$ scaling with time in the transient time. In Fig.~\ref{fig:qfi_lr}(a), we illustrate such behavior for fixed $\alpha=1.5$, where $\tilde{\beta}$ increases from $\tilde{\beta}_1=3.948$ to $\tilde{\beta}(\mathcal{Z}=N/2, \alpha=1.5)=3.982$, exhibiting the benefit of increasing  the range of interactions.

{\it Effect of fall-off rates.} A complementary way to quantify the role of long-range interactions is to fix the coordination number \(\mathcal{Z}\) and vary the decay exponent \(\alpha\), which controls the spatial profile of the couplings and  is experimentally tunable parameter in platforms  where power-law interactions can be engineered over a wide range. In this setting, reducing \(\alpha\) (i.e., making the interactions more long-ranged) systematically enhances the transient-time scaling of the QFI. For example,
%The increase in scaling of QFI in transient time is also obtained by decreasing fall-off rates $\alpha$, i.e., increasing long-range interactions. As shown in Fig.~\ref{fig:qfi_lr}(b) 
for $\mathcal{Z}=4$, the scaling exponent $\tilde{\beta}(\mathcal{Z}=4, \alpha=5.0)=3.951$ to $\tilde{\beta}(\mathcal{Z}=4, \alpha=0.8)=3.974$, thereby expressing the advantage of long-range interactions (see also  Fig.~\ref{fig:qfi_lr}(b)).

%Therefore, 
{\bf Note.} The above analysis clearly indicates that in the presence of long-range interactions, the scaling exponent increases $\tilde{\beta}(\mathcal{Z}, \alpha)>\tilde{\beta}_1$ with both the increase in coordination number $\mathcal{Z}$, and decrease in the fall-off exponent $\alpha$, although  the magnitude of QFI $\mathcal{F}_h^{nH}$ may decrease slightly. 

\begin{figure}
    \includegraphics[width=\linewidth]{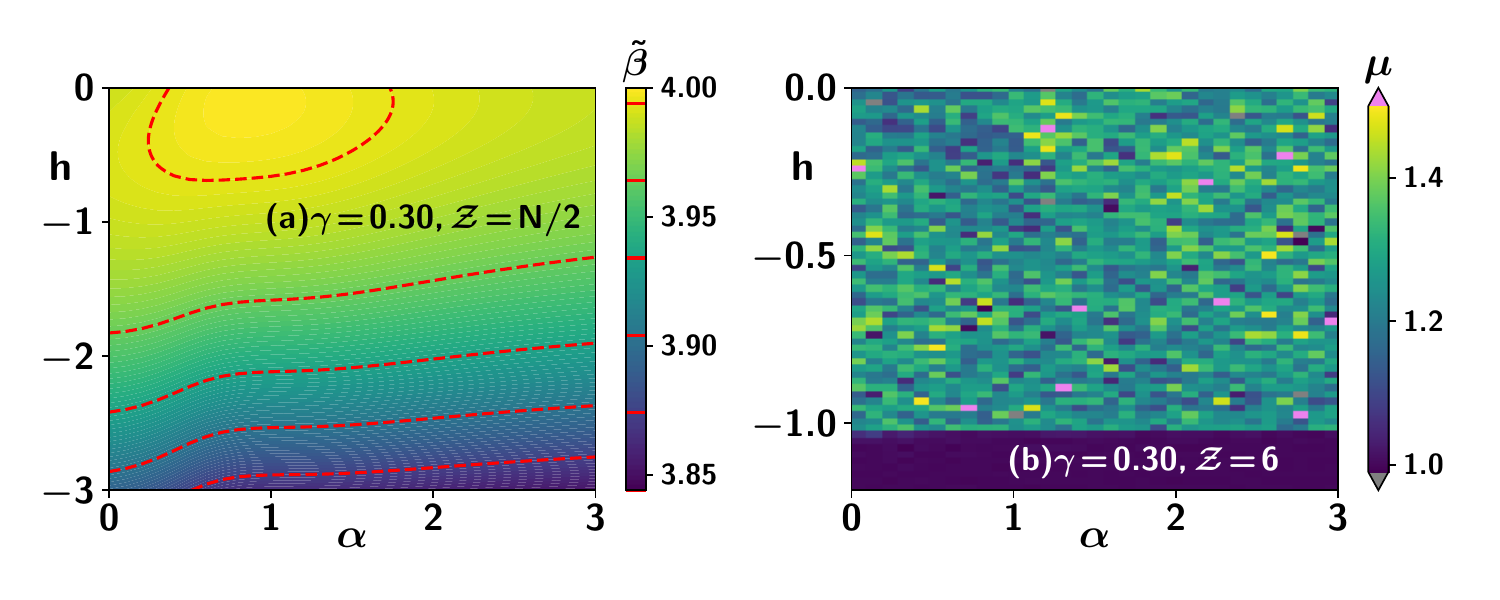}
    \caption{(a) Contourplot of scaling exponent $\tilde{\beta}$ of QFI in transient time in the  \((\alpha, h)\)-plane for a fixed \(\mathcal{Z}=N/2\).  (b) Map plot of system-size scaling exponent, $\mu$, by varying fall-off rates $\alpha$ (abscissa) and magnetic field $h$ (ordinate) with \(\mathcal{Z}=6\).  Here, in (a),  \(N=1024\),   $t=200$ in (b) and in both (a) and (b), \(\gamma =0.3\).  These plots indicate that the temporal and system-size scaling obtained from the dynamics is independent of phases.  All axes are dimensionless.}
    \label{fig:scaling}
\end{figure}

{\it Gain persists in broken and unbroken regimes.} Given the non-Hermitian nature of the model, it is natural to examine whether the observed scaling advantage depends on the underlying spectral phases, namely the broken and unbroken regimes. Our study reveals that the increase of the transient-time exponent \(\tilde{\beta}\) is obtained both in the broken and the unbroken regimes as illustrated in Fig.~\ref{fig:transient_t_scaling}(a) for the broken phase, and Fig.~\ref{fig:transient_t_scaling}(b) for the unbroken phase. In particular, \(\tilde{\beta}\) grows monotonically with both the increase in the range of interactions, for all values of the fall-off rates $\alpha$, and increases with decreasing $\alpha$ when $h\lesssim -1$. In contrast,  in the intermediate domain, $-1\lesssim  h<0$, the dependence of \(\tilde{\beta}\) with \(\alpha\) becomes non-monotonic (see Fig.~\ref{fig:scaling}(a)). In the long time limit, the dynamics of QFI is insensitive to these details and it universally approaches the standard quadratic scaling
with time. As we demonstrate in the next section, non-Hermitian dynamics is essential for achieving higher transient scaling exponents, even though the distinction between broken and unbroken phases does not qualitatively affect this enhancement.
%, i.e. $\mathcal{F}_h^{nH}(t)\sim t^\beta$, with $\beta\sim 2$ for all parameters of the $H^{iLR}$ model. 

% In the transient time, while the value of the QFI is similar or lower than in the Hermitian case, the non-Hermitian gives better scaling with time. The scaling coefficient with time increases with increase of the coordination number, the advantage over the Hermitian counterpart decreases, with both providing similar time sclaing in the transient time at large coordination numbers.

\begin{figure}
    \includegraphics[width=\linewidth]{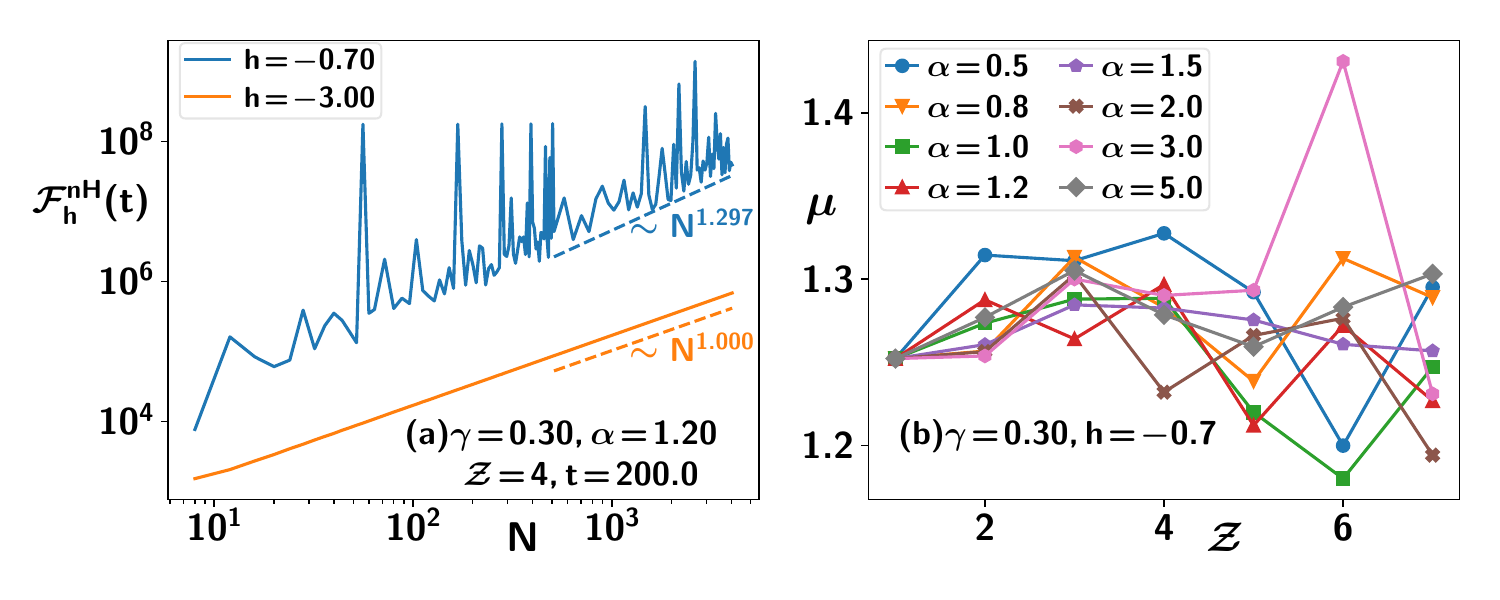}
    \caption{System-size scaling of QFI, given by $\mathcal{F}^{nh}_h\sim N^{\mu}$.  (a)  $\mathcal{F}^{nh}_h$ (ordinate) with system-size $N$ (abscissa) in the broken ($h=-0.5$) and the unbroken ($h=-3.0$) regimes, in the dynamics with long-range $iXY$ model. The scaling exponent is higher in the broken regime with $\mu>1$, giving a non-Hermitian advantage in the long-range system. (b) Scaling exponent $\mu$ (ordinate) against the coordination number $\mathcal{Z}$ (abscissa) for various fall-off rates $\alpha$. All other specifications are the same as in Fig. \ref{fig:qfi_lr} and all axes are dimensionless.}
    \label{fig:N_scaling}
\end{figure}

\subsection{Dynamical QFI scaling with system-size: Benefit of Non-Hermitian LR model }

The system-size \(N\) constitutes a key metrological resource 
as larger probes generally provide better precision in metrology. 
In the conventional setting, the QFI scales at most linearly with system-size, 
%increases with increase in system-size, with
$\mathcal{F}\sim N$ corresponding to the SQL. In equilibrium protocols based on Hermitian Hamiltonians, enhanced scaling can arise near critical points, while in dynamical settings, improvements are typically linked to the presence of entanglement~\cite{Toth_pra20212}. For initially separable probes and \(k\)-body encoding Hamiltonians, the scaling is bounded as
$\mathcal{F}\sim N^\mu$, where $\mu\leq k$ \cite{Puig2025}; in particular, for sensing magnetic fields (i.e., $k=1$) with product probes, Hermitian models can only provide \(\mu \leq 1\). 
In contrast, our analysis of the long-range non-Hermitian model reveals a qualitatively different behavior, with a strong dependence on the underlying spectral phase.

%the classical limit. While rate of increase, has shown quantum benefit near phase transitions in equilibrium protocols (where parameter is encoded by taking the system to the ground state of the model), and role of entanglement in the dynamical framework~\cite{Toth_pra20212}. In the absence of entanglement in the initial quantum probe, QFI is known to scale as $\mathcal{F}\sim N^\mu$, where $\mu\leq k$ for encoding of $k$-body Hamiltonian interaction~\cite{Puig2025}. Therefore, for magnetic field sensing with initial product states, Hermitian models give $\mu\leq 1$ limit. Here we study the effect of the long-range interactions in the non-Hermitian metrology in both the unbroken and the broken phases. 

{\it Broken regime.} In the broken phase, the model possesses a complex energy spectrum and eigenvectors which break the $\mathcal{RT}$-symmetry of the model. Therefore, one may expect that the dynamics governed by the Hamiltonian in the broken phase may exhibit genuinely non-Hermitian features that strongly impact metrological scaling and can be different 
%the properties of dynamical state can differ
from the (pseudo-)Hermitian model \cite{chu_prl_2020, Edvardsson_prb_2022, ding_prl_2023, xiao_prl_2024}. 
%In quantum metrology, such phases become sensitive to the underlying parameters, and was can surpass the scaling bounds given by the Hermitian model.

For the magnetic field sensing with initial product state, we observe that the dynamical QFI displays a nonmonotonic but overall superlinear growth with system-size for small \(\alpha (<2)\) and high \(\mathcal{Z} (\neq 1)\), thereby surpassing the Hermitian bound and highlighting also the significance of LR interactions. For example, $\mu\sim 1.3$ in $\mathcal{F}_h^{nH}\sim N^\mu$ along with fluctuations in the broken phase (see also Figs.~\ref{fig:N_scaling} and \ref{fig:scaling}). Note also that the nearest-neighbor non-Hermitian $\mathcal{RT}$-symmetric model also beats the $\mu\leq 1$ bound with encoding in the broken phase~\cite{agarwal2025}. 
%Such advantage is also seen in presence of long-range interactions. The QFI $\mathcal{F}_h^{nH}$ increase non-monotonically in the broken regime with the system-size $N$ (see Fig.~\ref{fig:N_scaling}(a) for $h=-0.7$). The overall increase of QFI with system-size is obtained with $\mu\sim 1.3$ in $\mathcal{F}_h^{nH}\sim N^\mu$ along with fluctuations in the broken phase. 
As depicted in Fig.~\ref{fig:N_scaling}(b), such quantum advantage of the non-Hermitian model persists across different inetraction ranges and decay exponents as well as for long evolution
%present at other values of coordination number $\mathcal{Z}$, as shown in Fig.~\ref{fig:N_scaling}(b) for various fall-off rates $\alpha$. 
% Interestingly, the exponent increases with smaller values of fall-off rates $\alpha$, at coordination number $\mathcal{Z}=2$. 
%Such increase in scaling exponent is seen 
%at all
times $t\gtrsim 200$.
%and various fall-off rates $\alpha$ for magnetic field $h$ in the broken regime, as shown in Fig.~\ref{fig:scaling}. 
This illustrates that the combination of non-Hermiticity and long-range interactions enables improved scaling beyond the SQL.
%the advantage of non-Hermiticity in long-range interactions for quantum sensing.

{\it Unbroken domain.} In stark contrast, in the unbroken regime where the spectrum remains real and the dynamics is effectively quasi-Hermitian, the QFI scales linearly with the system-size, i.e., $\mathcal{F}_h^{nH}\sim N$ for all parameters of the non-Hermitian $H^{iLR}$ model at all times.  While long-range interactions enhance overall sensitivity, the emergence of the symmetry-broken regime enables a genuine scaling advantage beyond Hermitian limits, highlighting the role of non-Hermitian physics as a resource for quantum sensing. Such a benefit will be analyzed more quantitatively in the next section.

%\section{Effect of long-range interactions on the non-Hermitian advantage over the Hermitian model}

\section{Genuine Non-Hermitian benefit in magnetometry: Hermitian vs non-Hermitian LR model}
\label{sec:nH_by_H}

Having established that long-range non-Hermitian interactions can enhance the QFI relative to their nearest-neighbor counterparts, it is natural to ask whether this gain is genuinely tied to non-Hermiticity, or whether a comparable improvement could already arise in an equivalent Hermitian long-range sensor. To isolate this distinction, we benchmark the present model 
%a Hermitian reference 
%we now compare the present model
with its Hermitian counterpart 
possessing the same interaction geometry and coupling scales.
% The previous section established that long-range couplings can significantly improve the sensing performance of the non-Hermitian sensor relative to the nearest-neighbor case. A more stringent question, however, is whether this improvement survives when benchmarked against a Hermitian model with identical interaction range and energy scales. We therefore now quantify the genuine metrological contribution of non-Hermiticity in the long-range setting.
%  The present long-range model provides a natural platform to examine how this advantage is modified when extended interactions are simultaneously present.

To quantify the genuine contribution of non-Hermiticity, we compare the QFI of the Hamiltonian $H^{iLR}$ with that of a Hermitian reference model obtained through the substitution $i\gamma \rightarrow \gamma$, while keeping the field strength $h$, the long-range coupling profile $J_r(\alpha)$, the coordination number $\mathcal{Z}$, and the initial state unchanged. In this manner, the geometric structure and overall interaction scale are preserved, such that any difference in sensing performance can be attributed directly to the non-Hermitian character of the dynamics. To isolate this effect, we monitor the time-averaged ratio between the non-Hermitian  and Hermitian sensors under otherwise identical conditions, which is defined as
\begin{eqnarray}
   \langle r^{nH}_{h} \rangle \equiv \Big\langle \frac{\mathcal{F}_{h}^{nH}(t)}{\mathcal{F}_{h}^{H}(t)}\Big\rangle
= \int_{t_0}^{t_1}  \frac{\mathcal{F}_{h}^{nH}(t)}{\mathcal{F}_{h}^{H}(t)} dt,
\label{eq:ratiotimeavg}
\end{eqnarray}
where the integration is performed over large times, i.e., \(t_0\) and \(t_1\) are sufficiently large (in simulation, we consider \(t_0=200\) and \(t_1 = 1000\)). Enhancements above unity, therefore, directly quantify the genuine contribution of non-Hermiticity. 

\begin{figure}
    \includegraphics[width=\linewidth]{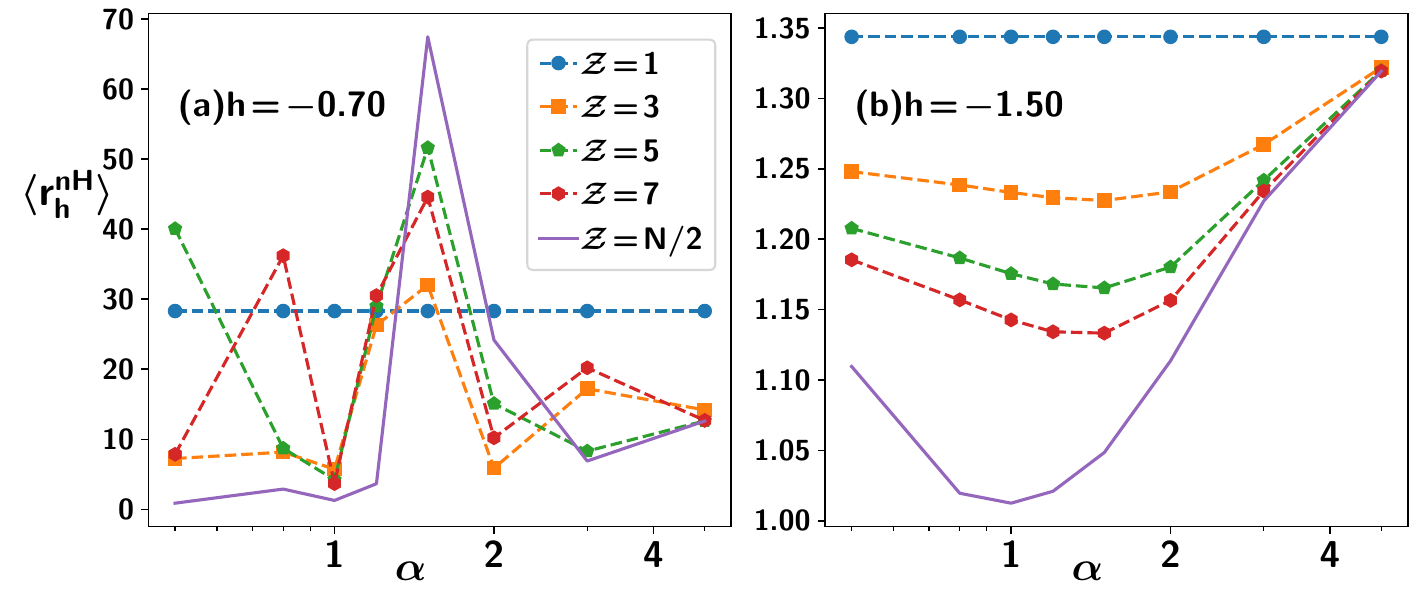}
    \caption{Time-average ratio of QFI in the non-Hermitian $iXY$ model with QFI in the Hermitian $XY$ model, defined in Eq. (\ref{eq:ratiotimeavg}). Precisely, \(\langle r^{nH}_{h} \rangle  \)  (ordinate) vs \(\alpha\) (abscissa). In (a) $h=-0.7$, and (b) $h=-1.50$, which are in the broken and the unbroken regimes of the $iXY$ model, respectively. Non-Hermitian models always provide benefits as the ratio is always above unity. In the broken regime (a), there are intermediate \(\alpha\)  values for which more precision can be obtained than the ones with low and high \(\alpha\) values while in the unbroken regime (b), nonmonotonic but smooth behavior of \(\langle r^{nH}_{h} \rangle  \) energes.   Here the average of the ratio is taken from $t=200$ to $1000$ for a fixed system-size $N=1024$. All axes are dimensionless.}
    \label{fig:qfi_nH_by_H}
\end{figure}

Two independent mechanisms can influence this ratio: the non-Hermitian spectral response itself, and the redistribution of correlations induced by long-range couplings. It is, therefore, instructive to first identify the nearest-neighbor limit, $\mathcal{Z}=1$, as a reference point. Even in this strictly local setting, the ratio already exceeds unity in both spectral regimes, demonstrating that the enhancement is intrinsic to non-Hermitian dynamics and does not require extended interactions. The dashed (blue) horizontal lines in Fig.~\ref{fig:qfi_nH_by_H} (a) and (b) represent this benchmark.

The magnitude of this enhancement, however, depends strongly on whether the system lies in the broken or unbroken phase. In the broken regime (Fig.~\ref{fig:qfi_nH_by_H}(a), $h < h_{e}$), the nearest-neighbor ratio is already of order $\mathcal{O}(10)$, indicating a pronounced gain over the Hermitian counterpart. This is consistent with the presence of complex quasiparticle energies and non-orthogonal eigenvectors, which together can strongly amplify dynamical susceptibility under normalized non-unitary evolution. By contrast, in the unbroken regime (Fig.~\ref{fig:qfi_nH_by_H}(b), $h > h_{e}$), where the quasiparticle spectrum remains real, the enhancement is far more modest and remains of order $\mathcal{O}(1)$. Thus, while non-Hermiticity improves sensing in both cases, the broken phase provides a substantially stronger metrological resource.

We now turn to the role of increasing interaction range. In the broken regime, the dependence on $\mathcal{Z}$ and $\alpha$ is highly nontrivial. For several values of $\alpha$, the ratio fluctuates around the nearest-neighbor benchmark, indicating that long-range connectivity alone does not guarantee additional improvement. Indeed, both small-$\alpha$ and large-$\alpha$ regions display irregular behavior with no universal monotonic trend. Interestingly, the fully connected limit $\mathcal{Z}=N/2$ often yields one of the smallest gains away from the optimal region, showing that maximal collectivity is not, by itself, sufficient. This possibly indicates an optimal regime in which algebraic interactions and non-Hermitian amplification act constructively to boost parameter sensitivity.

A qualitatively different behavior emerges near $\alpha \approx 1.5$, where the ratio increases systematically with coordination number. Here the curves become approximately ordered in $\mathcal{Z}$, and the fully connected model attains the largest enhancement in the panel, approaching nearly twice the nearest-neighbor benchmark. This indicates a cooperative interplay between long-range transport and non-Hermitian amplification. In other words, neither strictly local couplings nor completely uniform connectivity appears optimal; rather, an intermediate algebraic decay exponent provides the most favorable balance for parameter encoding.

The unbroken regime exhibits the opposite tendency. For $h=-1.50$, increasing $\mathcal{Z}$ generally suppresses the ratio across essentially the entire range of $\alpha$, as shown in Fig.~\ref{fig:qfi_nH_by_H}(b). The curves become progressively ordered with coordination number, revealing an almost monotonic decrease of the non-Hermitian advantage as the interaction graph becomes more collective. Since the spectrum remains real, the system lacks the mode-selective growth mechanisms characteristic of the broken phase, and extended couplings instead tend to homogenize the response, thereby reducing the relative gain over the Hermitian benchmark. 

Despite this overall suppression, the $\alpha$ dependence remains nontrivial. In particular, a pronounced minimum develops around $\alpha \sim 1$, most clearly for larger values of $\mathcal{Z}$, after which the ratio rises again toward larger $\alpha$. If long-range couplings were simply detrimental, one would expect a smooth monotonic interpolation between the collective and local limits. The observed dip instead suggests a competition between algebraic information spreading, finite connectivity, and the structure of the non-Hermitian eigenmodes. Such intermediate-$\alpha$ anomalies are reminiscent of crossover behavior known in long-range interacting systems, where transport fronts and correlation propagation reorganize near comparable exponents \cite{Hauke_tagliocozzo_long_range_prl}.

Taken together, these results demonstrate that long-range interactions do not play a universal role in non-Hermitian metrology. In the broken regime, they can significantly reinforce the already present non-Hermitian enhancement, but only within selected windows of the decay exponent. In the unbroken regime they instead reduce the advantage in a largely systematic manner. The optimal sensing architecture is therefore governed not by non-Hermiticity or long-range coupling separately, but by their combined interplay across distinct spectral phases.

\section{No-LR Gain over Hermitian in the stationary setting}
\label{sec:disadvantage_equilibrium}

We now turn to the question of whether long-range interactions can enhance magnetic-field sensing compared with the NN model, when the parameter is encoded in the stationary properties of the non-Hermitian Hamiltonian, rather than via dynamical evolution. We find the answer negatively. 
%in a \emph{static} non-Hermitian protocol. 
%In contrast to the dynamical setting discussed earlier, the parameter is here encoded in a stationary state of the Hamiltonian. 
This answer allows us to isolate the role of spectral singularities from that of time-dependent amplification mechanisms that drive the benefits observed in dynamical protocols.

{\bf Proposition.}    The scaling of QFI obtained with LR non-Hermitian model coincides with that of the NN model.
    
{\it Proof.} The gap-closing structure of the long-range non-Hermitian model $H^{iLR}$ follows directly from this dispersion relation in Eq.~(\ref{eq:dispersion}). In the thermodynamic limit, where $\phi_p$ becomes continuous, the imaginary part $J_p^{(I)}$ vanishes at the Brillouin zone boundaries $\phi_p=0$ and $\phi_p=\pi$. The condition $\epsilon(\phi_p)=0$, therefore, reduces to $h+J_p^{(R)}=0$. 
Using the Kac-normalized couplings, one obtains
\begin{equation}
    h_c^{(0)}=-1,
\end{equation}
and
\begin{equation}
    h_c^{(\pi)}
    =
    -\frac{\sum_{r=1}^{\mathcal Z}(-1)^r r^{-\alpha}}
    {\sum_{r=1}^{\mathcal Z}r^{-\alpha}}
    =
    1-\frac{2^{1-\alpha}H_{\lfloor \mathcal Z/2 \rfloor}^{(\alpha)}}{H_{\mathcal Z}^{(\alpha)}},
\end{equation}
where $H_n^{(\alpha)}=\sum_{r=1}^{n}r^{-\alpha}$ denotes the generalized harmonic number. 
Thus, the long-range model possesses two candidate critical fields associated with the momenta $\phi_p\to0$ and $\phi_p\to\pi$. We, however, stick to the critical point $h_c^{(0)} = -1$, throughout the paper.

The probe is taken as the dominant eigenstate of the LR non-Hermitian Hamiltonian. In the unbroken regime, this corresponds to the eigenstate with the lowest real eigenvalue, while in the broken region, where eigenvalues acquire imaginary components, the probe is identified with the eigenstate having the largest imaginary eigenvalue, which governs the normalized long-time behavior. To distinguish the resulting finite-size scaling from the dynamical case, we denote this stationary scaling exponent of the quantum Fisher information by $\mu_h^s$.

\begin{figure}
    \includegraphics[width=\linewidth]{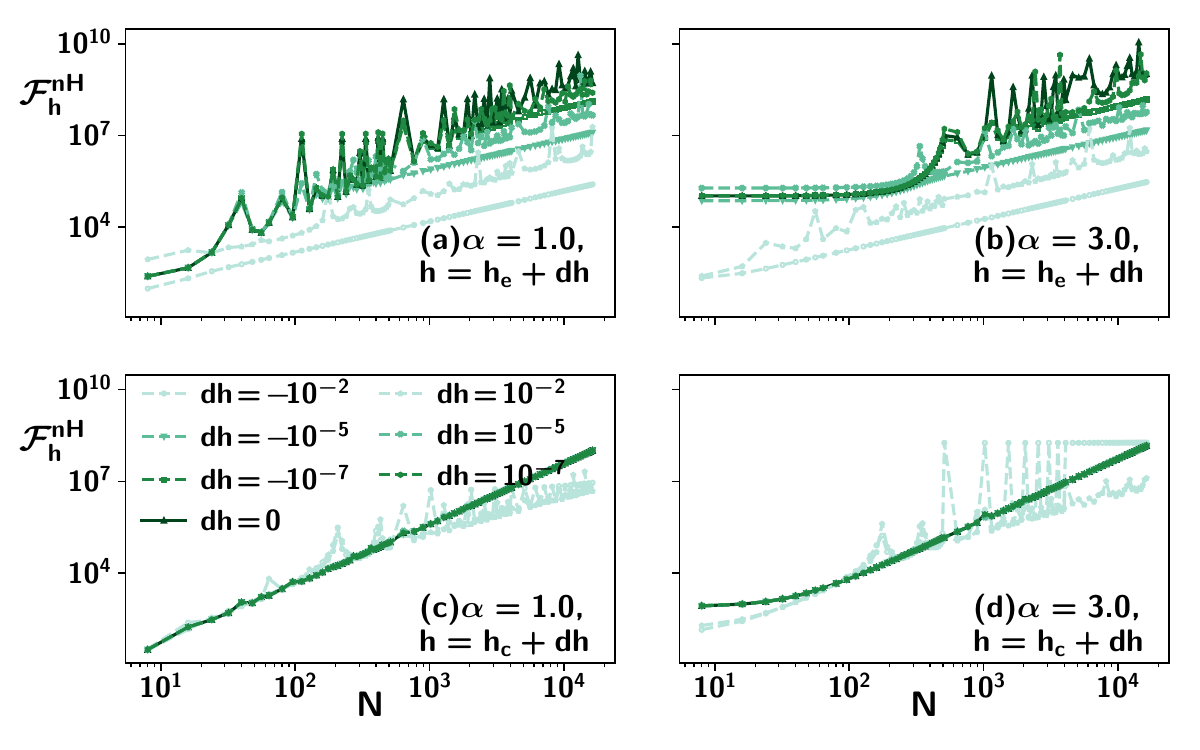}
    \caption{Role of next-nearest neighbor interactions ($\mathcal{Z}=2$) on QFI in magnetic field sensing. Specifically,  \(\mathcal{F}^{nH}_h\) (ordinate) of the eigenstate with minimum energy against \(N\) (abscissa) near the exceptional (\(h= h_e + dh\))  [(a)-(b)] and the critical points (\(h= h_c + dh\)) [(c)-(d)] for different fall-off parameter $\alpha$. $\gamma=0.5$ and \(N=1024\). All axes are dimensionless.}
    \label{fig:qfi_h_z2}
\end{figure}

\begin{figure}
    \includegraphics[width=\linewidth]{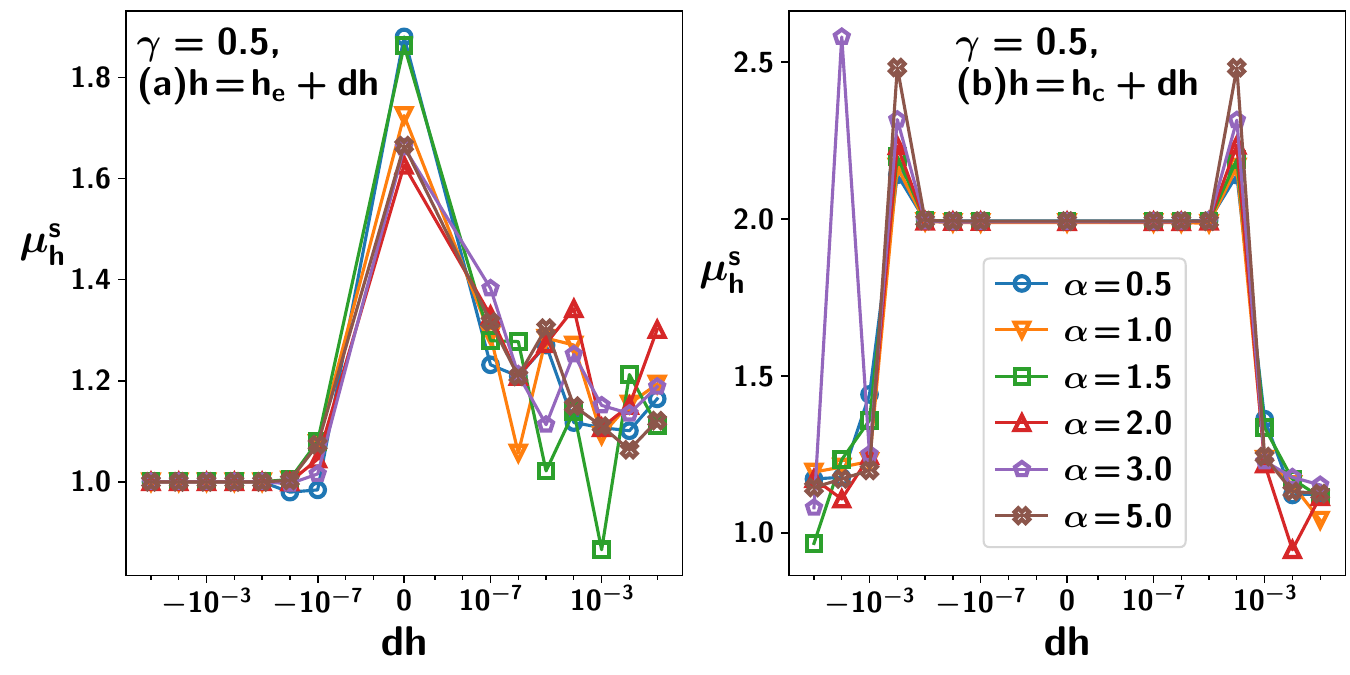}
    \caption{ Scaling \(\mu_h^s\) (ordinate) vs \(dh\) (abscissa) near the exceptional point (a) and critical point (b). Different lines with different symbols correspond to distinct \(\alpha \) values. 
    %Role of next-nearest neighbor ($\mathcal{Z}=2$) on the scaling of QFI in magnetic field sensing with the system-size, i.e., $\mathcal{F}_h^{nH}\sim N^{\mu_h^s}$. The scaling is seen near the (a) exceptional and the (b) critical point for different fall-off parameter $\alpha$. Here the imaginary anisotropy is 
    Here $\gamma=0.5$, \(\mathcal{Z}=2\) and the scaling coefficient is obtained for $N>10^3$. 
    All axes are dimensionless.}
    \label{fig:static_qfi_vs_h}
\end{figure}

\begin{figure}
    \includegraphics[width=\linewidth]{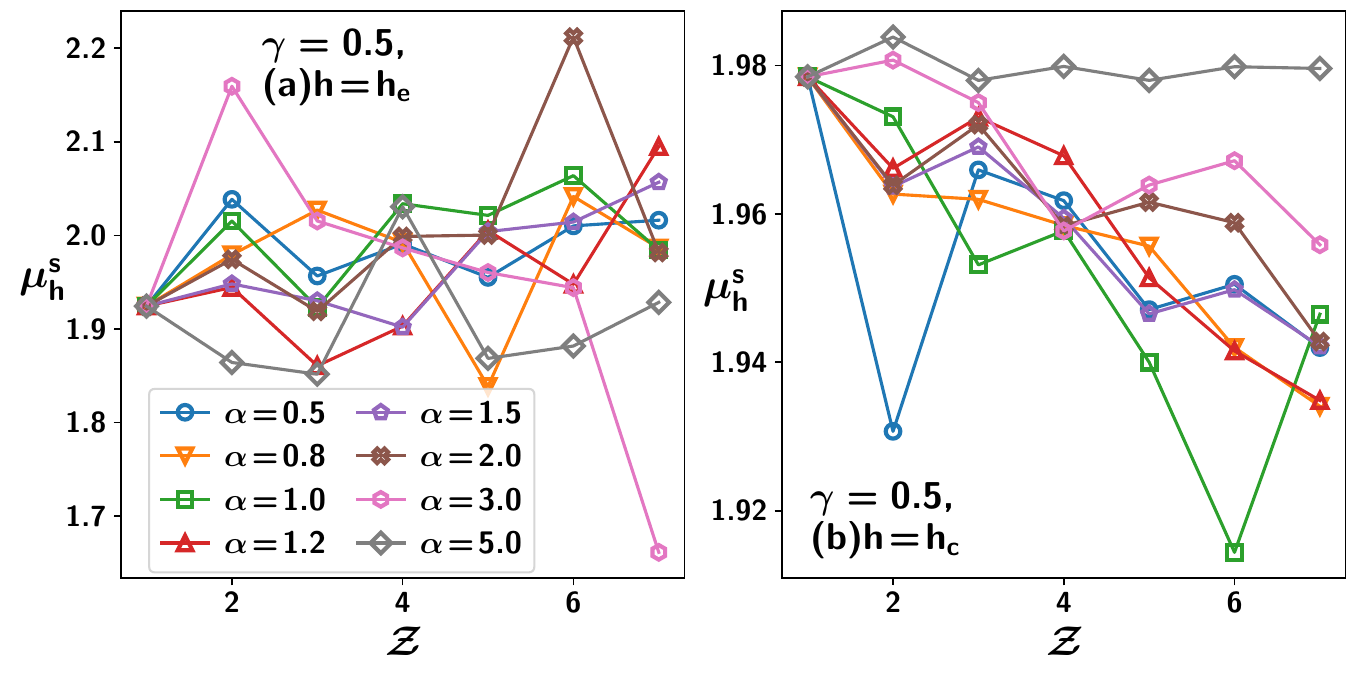}
    \caption{Impact of \(\mathcal{Z}\) on scaling  exponent of the stationary states for magnetic field sensing. By fixing \(h\) near the exceptional (a) and critical points (b), \(\mu_h^s\) vs \(\mathcal{Z}\) for various values of \(\alpha\). Especially near the exceptional point, there exist several \(\alpha\) values for which the scaling exponent is much higher than that obtained through the NN model. 
    %Role of coordination number $z$ on the scaling of QFI in magnetic field sensing with the system-size, i.e., $\mathcal{F}_h^{nH}\sim N^{\mu_h^s}$. The scaling is seen near the (a) exceptional and the (b) critical point for different fall-off parameter $\alpha$. Here the imaginary anisotropy is $\gamma=0.5$ and the scaling coefficient is obtained for $N>10^3$. 
    All axes are dimensionless.}
    \label{fig:static_qfi_vs_z}
\end{figure}

%The static scaling behavior is controlled by the proximity to 

We find that 
any enhancement of 
%spectral singularities. 
%The exponent 
$\mu_h^s$ 
%exhibits enhancement 
is confined only to a narrow window around the quantum critical point and the exceptional point, quantitatively characterized by $|dh|\lesssim10^{-3}$. Within this window, $\mu_h^s$ increases above its extensive value, exhibiting superextensive scaling, while outside it rapidly returns to $\mu_h^s\simeq1$, indicating the absence of collective sensitivity. This feature is observed for all fall-off exponents $\alpha$ considered, showing that long-range interactions do not generically extend the regime over which enhanced scaling occurs (see Fig.~\ref{fig:static_qfi_vs_h}).

Although the detailed profile of the enhancement depends on $\alpha$, the maximal value of $\mu_h^s$ attained near the singular points does not display a systematic dependence on the interaction range. Near the exceptional point, the extracted scaling exponent shows stronger fluctuations within the narrow $dh$ window, reflecting increased sensitivity to finite-size effects. Close to the quantum critical point, by contrast, the enhancement forms a comparatively stable plateau with $\mu_h^s$ remaining close to its maximal value across different $\alpha$. In both cases, however, long-range couplings modify the width and structure of the crossover but do not enhance the maximal scaling itself.

Further insight is obtained by varying the coordination number $\mathcal{Z}$. Increasing $\mathcal{Z}$ incorporates progressively more distant neighbors while preserving the overall interaction strength through Kac normalization. The resulting QFI scaling shows that $\mu_h^s$ fluctuates weakly with $\mathcal{Z}$ near both the exceptional and critical points, without exhibiting any monotonic trend or systematic improvement. The scaling exponent remains confined to the same range for all coordination numbers studied, demonstrating that increased connectivity does not translate into improved sensitivity (see Fig.~\ref{fig:static_qfi_vs_z}).

\hfill $\blacksquare$

These results establish that metrological enhancement in the long-range non-Hermitian chain is entirely governed by the singular spectral structure already present in the short-range (SR) model. Exceptional-point degeneracies and critical gap closing generate the dominant enhancement of the quantum Fisher information, while long-range interactions primarily affect nonuniversal features such as finite-size crossover behavior and the narrowness of the enhanced region in $dh$. They do not introduce additional divergences capable of improving the asymptotic scaling encoded in $\mu_h^s$.

We therefore conclude that, unlike in dynamical sensing protocols, LR interactions provide no genuine advantage for non-Hermitian sensing via stationary state. In this scenario, the achievable scaling of the QFI remains controlled by spectral singularities, and is essentially identical to that of the nearest-neighbor model, irrespective of the decay exponent $\alpha$ and the coordination number $\mathcal{Z}$.

%, the  quantum Fisher information near both the exceptional and the critical point remains governed by the same scaling behavior as in the nearest-neighbor model.

\section{Conclusion}
\label{sec:conclusion}

Recent technological advances call for a deeper understanding of many-body systems operating in realistic, open-system settings. When coupled to environments through dissipation, continuous monitoring, or post-selection, such systems are naturally described by non-Hermitian dynamics, leading to distinctive spectral features, including broken and unbroken phases separated by exceptional points. Concurrently, long-range (LR) interactions with power-law decay naturally arise in several experimental platforms, including trapped ions, polar molecules, and Rydberg atom arrays, where they profoundly influence the spread of correlations and entanglement. Here, we investigated the combined metrological implications of these two ingredients for magnetic-field sensing in a long-range non-Hermitian \(iXY\) spin chain, using the quantum Fisher information (QFI) as the figure of merit.

%Both non-Hermitian dynamics and long-range interactions have emerged as ubiquitous paradigms in quantum physics, expanding the landscape of accessible many-body phenomena. Non-Hermiticity naturally manifests in open quantum systems—driven by dissipation, continuous monitoring, or post-selection—giving rise to unique spectral features such as broken phase and exceptional points. Simultaneously, algebraically decaying long-range interactions arise naturally in quantum simulators, including trapped ions, polar molecules, and Rydberg atom arrays, where they fundamentally alter the propagation of correlations and entanglement. In this work, we explored the intersection of these two ubiquitous features to determine their combined potential for quantum metrology, specifically focusing on the precision bounds in estimation of a transverse magnetic field in a long-range non-Hermitian $iXY$ spin chain via quantum Fisher information. 

%Our results establish that the interplay of long-range connectivity and non-Hermiticity yields a genuine, robust sensing advantage in dynamical encoding protocols. 
In the dynamical encoding regime, we found that non-Hermiticity and long-range interactions cooperate to yield a genuine sensing advantage. Starting from a product initial state, we demonstrated that long-range couplings enhance the temporal scaling of the QFI in the transient regime. Crucially, the system-size scaling of the QFI surpasses conventional Hermitian bounds.
%($\mathcal{F}_h^{nH} \sim N^\mu$ with $\mu > 1$). 
Benchmarking the system against an equivalent Hermitian long-range model displays that this improvement originates from 
%we isolated this metrological enhancement to 
the broken $\mathcal{RT}$-symmetric phase, where non-Hermitian spectral amplification and non-orthogonal eigenstates with long-range interactions boost the parameter sensitivity. Interestingly, this advantage is not maximal in the model with purely uniform, infinite-range connectivity; instead, the relative enhancement peaks at intermediate decay exponents in the broken regime, pointing to an optimal balance between nonlocal information spreading and mode-selective amplification.

In stark contrast to the dynamical case, we established a rigorous no-go result when the target parameter is encoded in the stationary eigenstate near spectral singularities, such as the exceptional point or the quantum critical point. In this scenario, we show that the scaling of the QFI is entirely dictated by the underlying spectral singularities already present in the short-range model, thereby failing to provide any LR benefit. It clearly indicates that  LR interactions alone cannot improve metrological performance in the absence of dynamical amplification.  

%The finite-size scaling exponent of this QFI is governed by the nature of the singularity in energy spectrum already present in the nearest-neighbor model, remaining remarkably insensitive to variations in the coordination number $\mathcal{Z}$ or the decay exponent $\alpha$.

Our findings provide a comprehensive framework for understanding the role of interaction range in non-Hermitian quantum metrology and clarify that the benefits obtained from LR non-Hermitian systems are intrinsically dynamical.  These results offer insights for the design of advanced quantum sensors,
suggesting that experimental platforms such as driven-dissipative Rydberg arrays, trapped ions, and photonic networks can benefit from actively exploiting dynamical encoding schemes with long-range non-Hermitian many-body models.

%indicating that future experimental implementations, such as in driven-dissipative Rydberg atom arrays, trapped ions, or coupled photonic networks, should prioritize dynamical encoding schemes to fully exploit the cooperative advantages of long-range non-Hermitian physics.

\acknowledgements
We acknowledge the use of the cluster computing facility
at the Harish-Chandra Research Institute. KDA acknowledges “INFOSYS" scholarship for senior students. LGCL is funded by the European Union. Views and opinions expressed are, however, those of the author(s) only and do not necessarily reflect those of the European Union or the European Commission. Neither the European Union nor the granting authority can be held responsible for them. This project has been funded by the Caritro Foundation. This work was supported by the Provincia Autonoma di Trento, and Q@TN, the joint lab between the University of Trento, FBK—Fondazione Bruno Kessler, INFN—National Institute for Nuclear Physics, and CNR—National Research Council, Italy. This research was carried out and financed within the framework of the second Swiss Contribution MAPS (Grant No. 230870).  
\bibliography{ref}

\appendix

\section{Jordan-Wigner and Fourier transformation}
\label{app:jw_ft}

The long-range non-Hermitian $\mathcal{RT}$-symmetric model can be solved analytically using Jordan-Wigner transformation~\cite{barouch_pra_1970, barouch_pra_1971, LSM_main, santoro_ising_beginners_2020}, which maps the spin-$1/2$ (non-local) operators to the spinless fermionic creation and annihilation operators, given as
\begin{align}
    c_j &= \left( \prod_{m=1}^{j-1} \sigma_m^z \right) \sigma_j^- \quad ; \quad c_j^\dagger = \left( \prod_{m=1}^{j-1} \sigma_m^z \right) \sigma_j^+, \\
    \sigma^+_j &=  c_j^\dagger \prod_{m=1}^{j-1}(2 c^\dagger_m c_m -1) \ ;\ \sigma^-_j = i c_n\prod_{m=1}^{j-1}(2 c^\dagger_m c_m-1), \nonumber
\end{align}
where $\sigma_j^\pm = \frac{1}{2}(\sigma_j^x \pm i\sigma_j^y)$ are spin-ladder operators and $\sigma_j^z = 2c_j^\dagger c_j - 1$. The product of $\sigma^z$'s makes the transformation non-local, and gives the Jordan-Wigner string $Z_{j+1}^{j+r-1}$ between interacting sites $j$ and $j+r$ in $H^{iLR}$, i.e., $c_j c^\dagger_{j+r} = \sigma^-_jZ_{j+1}^{j+r-1}\sigma^+_{j+r}$ and similarly for other terms. In the case of boundary terms, for example, $\sigma^x_N\sigma^x_1$, an overall factor $\mathcal{K}=-\prod_{m=1}^{N}\sigma^z_m$. Since the Hamiltonian $H^{iLR}$ is $\mathbb{Z}_2$ symmetric in odd and even (parity) numbers of spin-up in the computational basis, $\mathcal{K}\equiv 1$ in the odd parity sector, whereas $\mathcal{K}\equiv -1$ in the even parity sector. Since the initial state in the dynamics is given by $\ket{\Psi(0)}=\ket{0}^{\otimes N}=\bigoplus_{p=1}^{N/2-1}\ket{0}_p$, which is in the even parity sector, the periodic boundary condition in spin-$1/2$ lattice, is transformed to anti-periodic boundary condition in the spinless fermionic picture, i.e., $c^{(\dagger)}_{N+k}\equiv c^{(\dagger)}_k$. In the static case as well, the eigenvector corresponding to the largest imaginary eigenvalue in the broken phase, and the lowest real eigenvalue in the unbroken phase, also belongs to the even parity sector for even system-sizes $N$, and we taken even system-size throughout the work. Therefore, in the fermionic picture, $H^{iLR}$ can be expressed as
\begin{align}
    H^{iLR} &= \sum_{j=1}^N \sum_{r=1}^{\mathcal{Z}} \frac{J_r(\alpha)}{2} \left[ (c_j^\dagger c_{j+r} + c_{j+r}^\dagger c_j) \right. \nonumber \\
    & \left. + i\gamma (c_j^\dagger c_{j+r}^\dagger - c_j c_{j+r}) \right] + \sum_{j=1}^N h \left(c_j^\dagger c_j - \frac{1}{2}\right)
\end{align}

Since the system is translationally invariant, the system can be further solved via Fourier transformation. Specifically moving to momentum space using the discrete Fourier transform, 
\begin{equation}
    c_j = \frac{1}{\sqrt{N}} \sum_p e^{i\phi_p j} c_p \ ;\ c_j^\dagger = \frac{1}{\sqrt{N}} \sum_p e^{-i\phi_p j} c_p^\dagger    
\end{equation}
where the anti-periodic boundary conditions for the fermions quantize the quasi-momenta as $\phi_p = \frac{(2p-1)\pi}{N}$, with $p$ from $-\frac{N}{2}+1$ to $\frac{N}{2}-1$ for even $N$. Therefore, the total momentum-space Hamiltonian becomes
\begin{align}
    H^{iLR} = \sum_{p=1}^{N/2-1} \left[ (h + J_p^{(R)}) \left( c_p^\dagger c_p + c_{-p}^\dagger c_{-p} \right) \right. \nonumber\\
    \left.+ \gamma J_p^{(I)} (c_p^\dagger c_{-p}^\dagger + c_p c_{-p}) \right] - hN/2,
\end{align}
and the Hamiltonian $H^{iLR}$ becomes block diagonal in the basis  $\{\ket{0}_p, c_p^\dagger c_{-p}^\dagger\ket{0}_p\}$ for $p=1$ to $\frac{N}{2}-1$, i.e., $H^{iLR}=\bigoplus_{p=1}^{N/2-1}\mathcal{H}_p$, with $\mathcal{H}_p$ as given in Eq.~(\ref{eq:ksea_P}) of the main text.

\section{Numerical values of the exceptional points}
\label{app:ep_points}

As emphasized in the main text, unlike the NN \(iXY\) model, the exceptional points cannot be obtained analytically for LR \(iXY\) Hamiltonian given in Eq. (\ref{eq:Hamil}). As part of our metrological analysis, we numerically determine the exceptional point \(h_e\) over a broad range of interaction parameters. In Table \ref{tab:hep_values}, we present the extracted values of \(h_e\) for different coordination numbers, 
\(\mathcal{Z} = 2,3,\ldots 7,N/2\) and decay exponents in the range \(0.5 \leq \alpha \leq 5 \), by numerically solving the dispersion relation in Eq.~(\ref{eq:dispersion}) with $\epsilon^2(\phi_p, h_{e}\!-\!\delta h)>0$ for all momentum $\phi_p$ and $\epsilon^2(\phi_p, h_{e}\!+\!\delta h)<0$ for some $\phi_p$, upto resolution $\delta h\!=\!10^{-9}$. Here, \(\alpha =5\) effectively reproduces the nearest-neighbor limit, while \(\alpha < 2\) corresponds to the genuinely long-range interacting regime.

Along with the study of quantum sensing, our simulation also reveals the exceptional points $h_{e}$ by varying \(\alpha\) and \(\mathcal{Z}\). The Table \ref{tab:hep_values} lists the exceptional point with different values of \(\mathcal{Z}=2,3,\ldots,7, N/2\) and \(0.5 \leq \alpha \leq 5\) values. Note that \(\alpha =5\) corresponds to the NN case while \(\alpha <2\) represents the LR models. 

\begin{table*}[t]
\centering
\label{tab:hep_values}
\begin{tabular}{c | c c c c c c c c}
\hline \hline
$\mathcal{Z} \setminus \alpha$ & 0.5 & 0.8 & 1.0 & 1.2 & 1.5 & 2.0 & 3.0 & 5.0 \\
\hline
2 & -1.105059766 & -1.104765248 & -1.104708169 & -1.104765291 & -1.105060404 & -1.106050573 & -1.109181424 & -1.114837818 \\
3 & -1.098162709 & -1.096888880 & -1.096282325 & -1.095903942 & -1.095804063 & -1.096901360 & -1.102543488 & -1.113373926 \\
4 & -1.093774407 & -1.091526878 & -1.090322573 & -1.089427233 & -1.088769814 & -1.089684072 & -1.097497978 & -1.112655607 \\
5 & -1.090690201 & -1.087573485 & -1.085810438 & -1.084413030 & -1.083180574 & -1.083811680 & -1.093557849 & -1.112278469 \\
6 & -1.088380338 & -1.084501800 & -1.082234348 & -1.080373509 & -1.078593376 & -1.078912992 & -1.090409422 & -1.112071808 \\
7 & -1.086572277 & -1.082024921 & -1.079305298 & -1.077023021 & -1.074735490 & -1.074744700 & -1.087849748 & -1.111956147 \\
$N/2$ & -0.999015181 & -1.012883927 & -1.011714837 & -1.006344872 & -1.001465522 & -1.013451483 & -1.076405943 & -1.111815671 \\
\hline \hline
\end{tabular}
\caption{Exceptional points $h_{e}$ for various coordination numbers $\mathcal{Z}$ and fall-off rates $\alpha$ for non-Hermitian anisotropy $\gamma=0.5$. In contrast, for the case of only the nearest-neighbor interactions, i.e., $\mathcal{Z}\!=\!1$, $h_e(\mathcal{Z}\!=\!1, \alpha, \gamma)=\sqrt{1+\gamma^2}$, as there is no effect of $\alpha$, giving ${h_e\sim-1.118033989}$ for $\gamma=0.5$.}
\end{table*}

\section{Sensing the anisotropy strength}
\label{app:gamma_sensing}

For sensing imaginary anisotropy strength $\gamma$ of the $\mathcal{RT}$-symmetric $H^{iLR}$ model in Eq~(\ref{eq:Hamil}), the QFI $\mathcal{F}_{\gamma}^{nH}$ can be computed via Eq.~(\ref{eq:qfi_theta}) with $\theta\equiv\gamma$ in both the dynamical and the stationary frameworks.
%We evaluate this in both the dynamical and stationary frameworks, establishing the ultimate precision bounds for estimating the degree of non-Hermiticity in the syst

In the dynamical protocol, starting from the initial state $\ket{\psi(t\!=\!0)}_p=\ket{0}_p$, the QFI scales as $\mathcal{F}_{\gamma}^{nH}\sim N^{\mu_\gamma} t^{\eta}$. We define the temporal scaling exponents in the transient and long-time regimes as $\tilde{\eta}$ and $\eta$, respectively. At long times ($t \gg 200$), the QFI exhibits characteristic quadratic scaling ($\eta=2$). In the transient regime, the temporal scaling remains sub-quadratic ($\tilde{\eta}<2$); however, $\tilde{\eta}$ systematically increases as the coordination number $\mathcal{Z}$ grows and the decay exponent $\alpha$ decreases. This behavior qualitatively mirrors the trends observed for $\tilde{\beta}$ in Fig.~\ref{fig:transient_t_scaling}, albeit with distinct numerical values. The system-size scaling exponent $\mu_{\gamma}$ behaves similarly as well, yielding $\mu_{\gamma}\sim 1$ in the unbroken regime and $\mu_{\gamma}\sim 1.4$ in the broken phase across all fall-off rates. It is important to note that because the parameter is encoded via long-range interactions (involving up to $N/2$-body terms), this super-extensive scaling does not violate the fundamental bounds of the Hermitian limit, as the encoding Hamiltonian remains quadratic in the spinless fermionic representation.

To understand the genuine advantage of non-Hermiticity, we compare $\mathcal{F}_{\gamma}^{nH}$ to its Hermitian counterpart, $\mathcal{F}_{\gamma}^{H}$. 
%The reference Hermitian model is obtained by substituting $i\gamma \to \gamma$, corresponding to an extended long-range $XY$ Hamiltonian. 
We note that while $\mathcal{F}_{\gamma}^{nH}$ quantifies the precision in estimating the non-Hermitian dissipation, $\mathcal{F}_{\gamma}^{H}$ bounds the estimation of the coherent exchange anisotropy. Analogous to the magnetic field sensing, we observe a distinct non-Hermitian amplification: the time-averaged ratio $\langle r^{nH}_{\gamma} \rangle \equiv \big\langle \mathcal{F}_{\gamma}^{nH}(t) / \mathcal{F}_{\gamma}^{H}(t) \big\rangle$ strictly exceeds unity for all considered parameters. This enhancement is highly pronounced in the broken phase, where $\langle r^{nH}_{\gamma} \rangle \sim \mathcal{O}(10)$, whereas it remains modest in the unbroken regime with $\langle r^{nH}_{\gamma} \rangle \sim \mathcal{O}(1)$.

\begin{figure}
    \includegraphics[width=\linewidth]{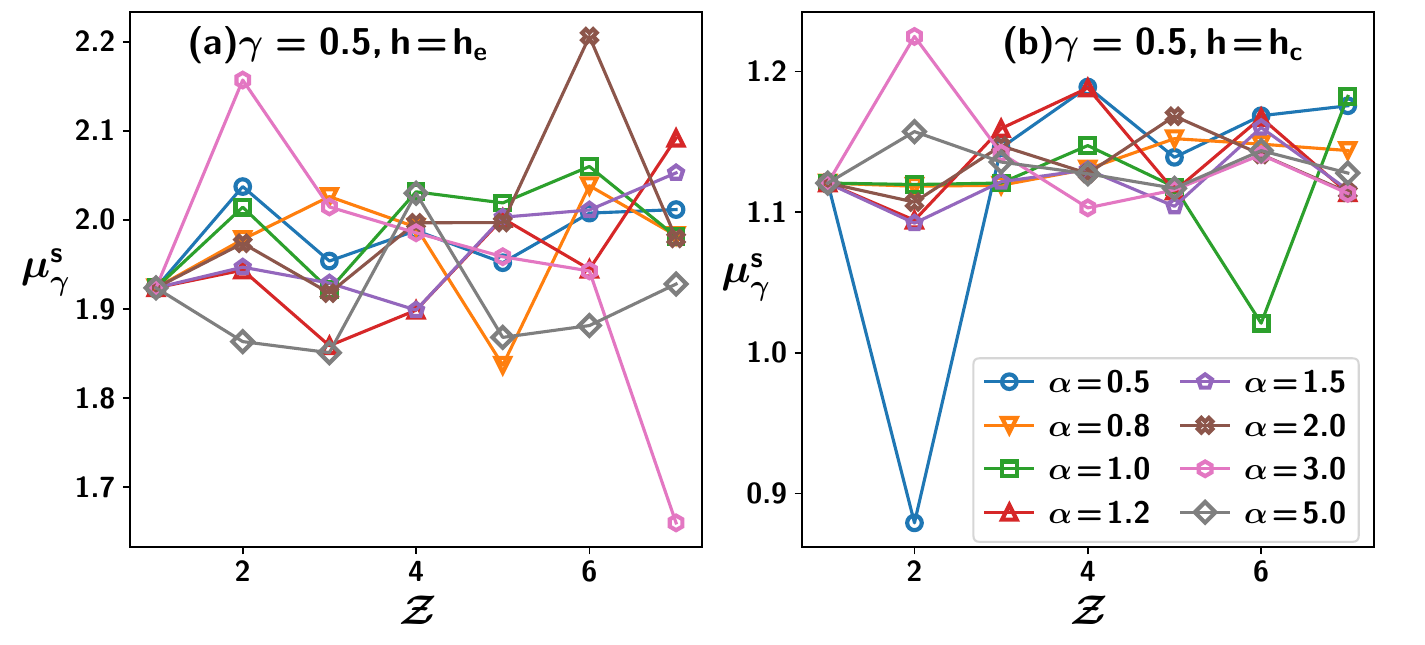}
    \caption{Scaling exponent \(\mu_{\gamma}^s\) of QFI for sensing non-Hermitian anisotropy $\gamma$ with system-size, i.e., $\mathcal{F}_\gamma^{nH}\sim N^{\mu_\gamma^s}$ at (a) the exceptional point, and (b) the critical point. This figure exhibits a similar system-size scaling behavior at exceptional point as observed in Fig.~\ref{fig:static_qfi_vs_z} under variation of the coordination number, with the only distinction being at the critical point with $\mu_\gamma^s\sim 1.1$. The estimated parameter here is \(\gamma\) rather than \(h\). All other specifications and interpretations are the same as in Fig.~\ref{fig:static_qfi_vs_z}. 
    All axes are dimensionless.}
    \label{fig:static_qfi_vs_z_gamma}
\end{figure}

    %Role of coordination number $z$ on the scaling of QFI in anisotropy sensing with the system-size, i.e., $\mathcal{F}_\gamma^{nH}\sim N^{\mu_\gamma^s}$. The scaling is seen near the (a) exceptional and the (b) critical point for different fall-off parameter $\alpha$. Here the imaginary anisotropy is $\gamma=0.5$ and the scaling coefficient is obtained for $N>10^3$. 
    
Finally, in the stationary framework, where the parameter is encoded within the dominant eigenstate of the long-range non-Hermitian Hamiltonian, the  QFI scales as $\mathcal{F}_\gamma^{nH}\sim N^{\mu_\gamma^s}$. As shown in Fig.~\ref{fig:static_qfi_vs_z_gamma}, we observe a clear quantum metrological advantage ($\mu_\gamma^s>1$) in the vicinity of both the critical and exceptional points. While this enhancement is modest near the critical point $h_c=-1$, yielding $\mu_\gamma^s\sim 1.1$, a strictly quadratic scaling ($\mu_\gamma^s = 2$) is attained exactly at the exceptional point, reflecting the divergent susceptibility of coalescing eigenstates. 

Therefore, both the estimation of the non-Hermitian interaction strength $\gamma$ and the magnetic field $h$ show that the advantage of long-range interactions in the transient time scaling is independent of the parameter being estimated, with long-range interactions increasing the transient-time scaling exponents universally. The system-size scaling in the dynamics with the system in the broken regime also demonstrates the advantage of non-Hermiticity with super-linear scaling, and $\sim O(10)$ amplification over the Hermitian counterpart, regardless of the parameter encoded. In the stationary framework, the exceptional and critical points provide different scalings for sensing the non-Hermiticity $\gamma$, distinguishing the nature of gap-closing between them. 

\end{document}